\documentstyle[seceq]{ptptex}

\input{epsf}

\def\qbar{{\bar q}}
\def\Im{\mathop{\rm{Im}}}

\def\g#1{\gamma^{#1}}

\def\Bt{{\tilde B}}

\newcommand{\beq}{\begin{eqnarray}}
\newcommand{\eeq}{\end{eqnarray}}

\newcommand{\del}{\partial}

\newcommand{\half}{\frac{1}{2}}
\newcommand{\thalf}{\textstyle \frac{1}{2}}
\newcommand{\hal}[1]{{#1 \over 2}}
\newcommand{\ket}{\rangle}
\newcommand{\bra}{\langle}
\newcommand{\bracket}[2]{\langle #1 | #2 \rangle}

\def\scdot{\! \cdot \!}
\def\tr{\,\hbox{tr}\,}

\def\gsim{\displaystyle\mathop{>}_{\sim}}

\newcommand{\Dslash}{{D\hspace{-7pt}/\, }}

\newcommand{\dslash}{\del \hspace{-6pt}/}
\newcommand{\kslash}{{k\hspace{-6pt}/}}
\newcommand{\pslash}{{p\hspace{-5pt}/}}

\newcommand{\gpNN}{g_{\pi N N}}
\newcommand{\gpNR}{g_{\pi N N^*}}
\newcommand{\gpRR}{g_{\pi N^* N^*}}
\newcommand{\geNR}{g_{\eta N N^*}}

\def\Journal#1#2#3#4{{#1} {\bf #2}, #3 (#4)}

\def\PRD{{\em Phys. Rev.} D}

\title{
Chiral Symmetry of Baryons
}

\author{
Daisuke {\sc JIDO}$^{1,}$\footnote{
e-mail address: Daisuke.Jido@hal.ific.uv.es}
Makoto {\sc OKA}$^{2,}$\footnote{
e-mail address: oka@th-phys.titech.ac.jp}
and
Atsushi {\sc HOSAKA}$^{3,}$\footnote{
e-mail address: hosaka@rcnp.osaka-u.ac.jp}
}

\inst{
$^1$ Instituto de F\'\i sica Corpuscular (IFIC), Centro Mixto
Universidad de Valencia-CSIC,
 Institutos de Investigaci\'{o}n de Peterna, Aptdo.
 Correos 2085, 46071, Valencia, Spain \\
$^2$ Department of Physics, Tokyo Institute of Technology,
 Meguro, Tokyo 152-8551 Japan\\
$^{3}$ Research Center for Nuclear Physics, Osaka University,
Ibaraki, 567-0047 Japan
 }

\recdate{
}

\abst{
We study chiral symmetry aspects of the positive and
negative  parity baryons by identifying
them with linear representations of the chiral group
$SU(N_{f}) \otimes SU(N_{f})$.
It is shown that there are two distinctive schemes: naive and mirror
assignments.
We construct linear sigma models for baryons 
in the two assignments and
examine their physical implications.
Then we investigate  properties of the naive and mirror nucleons
microscopically by using QCD interpolating fields.
Finally, we propose experiments to distinguish the two chiral
assignments for the nucleon.
}

\begin{document}

\maketitle

\section{Introduction}

Chiral symmetry plays a unique role in hadron physics.
In QCD, it is realized as the global
flavor symmetry for the right and left handed quarks
in the  light flavor sector.
The associated chiral group is then denoted as
$SU(N_{f})_{R} \otimes SU(N_{f})_{L}$, where $N_{f}$ is the number of light
flavors.
At low energies, however, it is spontaneously broken and only the
diagonal vector symmetry remains to manifest,
\beq
SU(N_{f})_{R} \otimes SU(N_{f})_{L}
\to
SU(N_{f})_{V} \, .
\eeq
The spontaneous breakdown is accompanied by the appearance of the
massless pseudo-scalar mesons, the pions and kaons,  as the
Nambu-Goldstone bosons~\cite{Nambu,Goldstone}.
In the real world the mesons acquire finite but small masses due to
small current quark masses.
Then the pions play the central role in strong
interaction physics at low energies.
Chiral symmetry has been proved to be useful in understanding dynamics
of low energy pions and nucleons as shown in the success of
the current algebra
based on conserved vector current (CVC) and partially conserved
axial-vector current (PCAC)~\cite{Weinbergbook}.

In spite of the success of such symmetry guided approaches, there
still remains a fundamental question;
what representations of the chiral group manifest in observed
particles.
In fact, such a question is not new.
Prior to QCD, much study was performed in the late
60th~\cite{Weinberg1,Gerstein}.
The reasons that we are now interested in this long standing
problem are:
(1) Much progress has been made in the fundamental theory of QCD.
(2) Relevant physics can now be accessible by experiments at the
present day.
(3) The problem of parity doubling, in particular, of the baryon
sector has not been considered much.
In this paper, we would like to focus on
the last issue, the chiral symmetry of baryons.

To start with, let us recall the analysis
by Weinberg about chiral representations for hadrons, in particular,
for baryons~\cite{Weinberg1}.
In a naive consideration of the relativistic nucleon filed,
it is natural to identify it with
the fundamental representations of the chiral group
$SU(2)_{R} \otimes SU(2)_{L}$~\footnote{Isospin values are
used in order to denote representations of the isospin $SU(2)$ group.
See, section 2
for definitions.}
\beq
\label{psichirep}
\psi \sim (\thalf, 0) \oplus (0, \thalf) \, .
\eeq
Such a linear field may be
coupled in a chiral invariant way with linear chiral mesons belonging
to a representation $(\thalf, \thalf) \sim (\vec \pi, \, \sigma)$.
This includes not only the pion but also the sigma meson $\sigma$,
the existence of which is still under controversy.
Then Weinberg motivated to eliminate such
poorly established particles and constructed
a theory only by well established particles, the pions,
which are linear
representations of the residual vector symmetry group~\cite{Weinberg2}.
The elimination is made by imposing  a constraint among
the linear chiral fields,
which necessarily lead to the non-linear realization of chiral symmetry.
In the non-linear realization
the lagrangian formulation serves a derivative expansion
for amplitudes and has become the
basis of the chiral perturbation theory~\cite{Gasser,Ecker,Pich}.
In the non-linear sigma model, since the spontaneous breaking
is assumed from the beginning, it does not seem easy to
discuss dynamical issues such as the phase transition of chiral symmetry.

\begin{figure}[tbp]
\centering
\epsfxsize = 12cm
\epsfbox{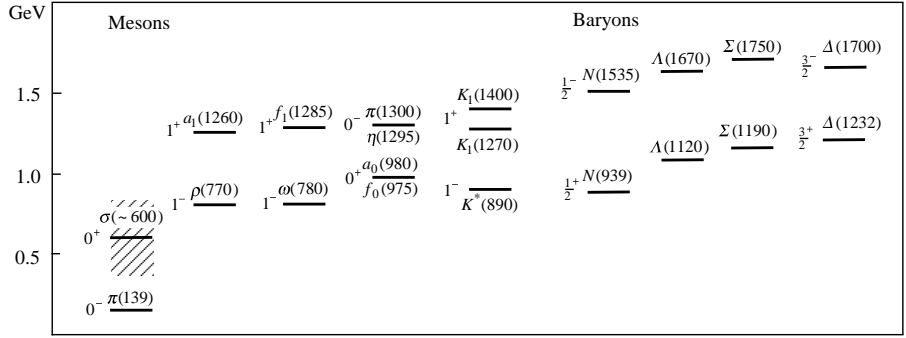}
\begin{minipage}{12cm}
   \caption{
   Mass splittings of positive and negative parity hadrons in various
   channels.
   Data are taken from the Particle Data Booklet~\cite{PDG}.
   The uncertain mass of sigma ($\sigma$) is hatched.
   \label{pnmass}}
\end{minipage}
\end{figure}

In this respect, linear sigma models are rather suited to the study
of dynamical aspects of chiral symmetry~\cite{Gell-Mann}.
For instance, it can describe the spontaneous breaking
within the model.
Also, linear representations introduce chiral partners having
different parities.
This provides a systematic view on hadrons
from chiral symmetry.
For instance, mass splittings of positive and negative parity hadrons
are generated by the spontaneous breaking of chiral symmetry and are
governed by a unique mass scale.
\ Furthermore,  they
become degenerate when chiral symmetry is restored.
These seem plausible since observed mass differences of positive and
negative parity hadrons are about 500 MeV independent of
particle channels as shown in Fig.~\ref{pnmass}.

Now in QCD, the quark field $q$ belongs to a linear representation of
the chiral group as
in Eq. (\ref{psichirep}).
Then the chiral mesons  are naively formed as quark bilinears:
\beq
(  i \bar q \gamma_{5} \vec \tau q, \; \bar q q) \sim
(\vec \pi\, , \; \sigma) \sim (\thalf, \thalf) \, .
\eeq
We observe that the scalar $\sigma$ and pseudoscalar $\vec \pi$ are
chiral partners among themselves.
Similarly, we can construct the $(1,0) \oplus (0, 1)$ representation for
the vector ($\rho$) and axial-vector ($a_{1}$) mesons.

Then what is the chiral representation of the nucleon?
Or, what is the chiral partner of the ground state nucleon?
This is not a trivial question, since the nucleon is a composite of
three quarks and is subject to strong interactions with the pions.
The simplest possibility has been already given in (\ref{psichirep});
four components of the nucleon
form a chiral representation among themselves, where
it is not necessary to introduce negative parity nucleons as a chiral
partner.

We can now list the relevant questions as follows:
\begin{itemize}
	\item
	Is it possible to consider larger representations
	which include both positive and negative parity nucleons and relate
	them by chiral symmetry?

	\item
	If so, what are possible representations
	and what dynamical consequences are expected?

	\item
    What is the quark (QCD) origin of such realizations?

    \item
    Are there any experimental approaches to inform the chiral symmetry
    of baryons?

\end{itemize}
These are the topics we would like to study in this article.

Much of the material treated here is based on our former and currently
progressing works~\cite{Jido1,Nemoto,Jido2,Jido3,Jido4,Kim}.
In fact, stimulated by the similar motivations,
Lee~\cite{Lee2} and later
DeTar and Kunihiro~\cite{DeTar}
previously constructed a theory
in which positive and negative parity nucleons behave as members of a
larger chiral representation.
What we will explore here is then to clarify the meaning of
their work from a fundamental point of view.
A somewhat unusual way of introducing the two nucleons in their
original works is now reformulated
using group theoretical languages.

In section 2,
after a  basic introduction for chiral symmetry and its
representations, we discuss group theoretical aspects of the
two chiral representations for positive
and negative parity nucleons.
In section 3, we study linear sigma models in the two assignments
and discuss phenomenological consequences.
Several comments are made for theoretical aspects of the models.
A microscopic origin of these two realizations is discussed in
section 4, where various baryon interpolating fields are investigated.
Since at the present time
the theoretical understanding has reached only to a
limited extent, we propose experiments
which will help discussing the two assignments.
For this, 
in section 5, we calculate reaction cross sections for the pion and
eta productions at the threshold region.
Final section is devoted to the summary of the present work.

\section{Representations of the Chiral Group for Baryons}

Here we discuss linear representations of the chiral symmetry group
for baryons  starting from chiral symmetry for quarks.  Finally
we shall reach two possibilities to assign the chiral symmetry to two
kinds of baryons, which makes different consequences of
phenomenological properties of baryons.

\subsection{Chiral Symmetry in QCD}
QCD Lagrangian is given in a very simple form:
\begin{equation}
{\cal L}_{\rm QCD} = \sum_f \bar{q}_f (i \Dslash - m_f ) q_f - {1
\over 4} F_{\mu\nu}^a F_a^{\mu\nu} \ ,
\end{equation}
where $q_f (f=u,d,s,c,b,t)$ are fields of quarks with mass $m_f$,
$F_{\mu\nu}^a$ denotes the field tensor of gluons with their color
index $a$. For the light quarks ($f = u$, $d$ and $s$), there is an
approximate symmetry independent of the chirality of the quarks, that
is called chiral symmetry. The definition of chiral symmetry with
$N_{f}$ light flavor is given
by flavor transformations on the right and left handed
components (of the Lorentz group):
\beq
q_{r} \to e^{i {\vec  t} \cdot \vec r} \psi_{r} \, , \; \; \;
q_{l} \to e^{i \vec t \cdot \vec l} \psi_{l} \, ,
\label{phtrsu}
\eeq
where $\vec t$ is a vector of $N_{f}^2 - 1$ generators of $SU(N_{f})$, and
$\vec r$ (and $\vec l$) are $N_{f}^2 -1$ transformation parameters.
The right and left handed components
are projected out from a
four component quark field $q$ by
\beq
q &=& q_{r} + q_{l} \, ,
\label{psisumrl}  \\
q_{r} &=& \frac{1+ \gamma_{5}}{2} q \, , \; \; \;
q_{l} \; =\;  \frac{1- \gamma_{5}}{2} q \, .
\label{psirandl}
\eeq
The generators satisfy the following commutation relations and
normalization:
\beq
 \left[ t^{a}, t^{b} \right] = i f_{abc} t^c \, , \; \; \;
\tr \left( t^{a} t^{b} \right) = \half \delta_{ab}\, ,
\label{ttcomm}
\eeq
where $f_{abc}$ are the structure constants of $SU(N_{f})$.

In the following discussions, we deal mostly with the case $N_{f} = 2$
for isospin $SU(2)$ symmetry.
Thus the generators are given by
\beq
t^a = \frac{\tau^a}{2} \, ,
\eeq
where $\tau^a$ are the Pauli matrices, and the structure constants
are $f_{abc} = \epsilon_{abc}$.

In the quantum theory,  transformations of the field
$q$ are generated by the charge operators.
For instance, the right handed field is transformed by the right
chiral charges $Q^a_{R}$,
\beq
\label{psiQRtr}
q_{r} \to
g_{R}^{} q_{r} g_{R}^\dagger\, ,  \; \; \;  g_{R}
\equiv \exp(iQ_{R}^a r_{a})
\, .
\eeq
The charges are defined by the Noether currents:
\beq
Q^a_R(t) = \int d^3 x \ J_{R0}^a(x) =
\int d^3 x q_{r}^{\dagger} t^{a} q_{r}  \, .
\eeq
Using the equal time commutation relations of the field $q$
we obtain the transformation rules of chiral symmetry
transformations as
\beq
i \left[ Q_{R}^a , q_{r} \right] &=& - i t^a q_{r} \, , \nonumber \\
\label{Qpsirlcomm}
i \left[ Q_{L}^a , q_{l} \right] &=& - i t^a q_{l} \, , \\
i \left[ Q_{L}^a , q_{r} \right]
&=& i \left[ Q_{R}^a , q_{l} \right] \; = \; 0 \, .
\nonumber
\eeq
Furthermore  we find the
commutation relations among the chiral charges,
\beq
 \left[ Q_{R}^a , Q_{R}^b \right] &=&
 i \epsilon_{abc} Q_{R}^c \, , \nonumber \\
 \left[ Q_{L}^a , Q_{L}^b \right] &=&
 i \epsilon_{abc} Q_{L}^c \, ,
\label{commQRL} \\
 \left[ Q_{R}^a , Q_{L}^b \right] &=& 0 \, .  \nonumber
\eeq
These are the basic commutation relations of the
chiral group $SU(2)_{R} \otimes SU(2)_{L}$.

It is worth emphasizing that the subscripts $R$ and $L$ are introduced
here in order to distinguish the symmetry groups assigned to
$q_{r}$ and $q_{l}$;  we could have adopted other labels
such as $A$ and $B$, etc.  The important point is that different
internal symmetries ($SU(2)$ and $SU(2)$) can be assigned to the right
and left handed fermions, independently.  
This is the reason that we use small letters $r$ and $l$ 
for the Lorentz group and large letters $R$ and $L$ for the chiral 
group.   

The commutation relations of (\ref{commQRL}) may be equivalently
expressed by the vector and axial charge operators,
\beq
Q_{V}^a = Q_{R}^a + Q_{L}^a \, , \; \; \;
Q_{A}^a = Q_{R}^a - Q_{L}^a \, ,
\label{defQVA}
\eeq
which satisfy
\beq
\left[ Q_{V}^a , Q_{V}^b \right] &=& i \epsilon_{abc} Q_{V}^c \, ,
\nonumber \\
\left[ Q_{V}^a , Q_{A}^b \right] &=& i \epsilon_{abc} Q_{A}^c \, ,
\label{commQVA} \\
\left[ Q_{A}^a , Q_{A}^b \right] &=& i \epsilon_{abc} Q_{V}^c \, .
\eeq
In this $VA$ representation, the right and left handed
fermion fields are transformed as
\beq
i \left[ Q_{V}^a , q_{r} \right] = - i t^a q_{r}\, ,
\; \; & & \; \;
i \left[ Q_{V}^a , q_{l} \right] = - i t^a q_{l}\, ,
\nonumber \\
i \left[ Q_{A}^a , q_{r} \right] = - i t^a q_{r}\, ,
\; \; & & \; \;
i \left[ Q_{A}^a , q_{l} \right] = + i t^a q_{l}\, .
\label{VAtransrl}
\eeq
In terms of a four component representation, these are combined into
\beq
i \left[ Q_{V}^a , q \right] = - i t^a q\, , \; \; \;
i \left[ Q_{A}^a , q \right] = - i \gamma_{5} t^a q\, , 
\label{VAtranspsi}
\eeq
where the $\gamma_{5}$ matrix is diagonal in the chiral (Weyl) 
representations:
\beq
\gamma_{5} = 
\left(
\begin{array}{c c}
    1 & 0 \\
    0 & -1
\end{array}
\right)
\, .
\eeq

Now we define a notation specifying representations of the chiral group
$SU(2)_R \otimes SU(2)_L$ as
\beq
\label{defrep}
(I_{R} , I_{L}) \, .
\eeq
where $I_R$ ($I_L$) denotes the isospin value of the representation of
the group $SU(2)_R$ ($SU(2)_L$). In this notation, the quark field is
expressed group theoretically as a direct sum of the fundamental
representation of the chiral group:
\beq
q = q_{r} + q_{l} \sim (\thalf,0) \oplus (0,\thalf) \, .
\label{psirlrep}
\eeq
The reason that the quark field 
is more than an irreducible representation
is that it is an eigenstate of a parity transformation.

%
%
%

\subsection{Representations for one baryon}
After defining chiral symmetry on the quark fields, 
we construct linear representations for baryons of chiral symmetry
from the quark fields.
One of the simplest way is to make direct
product of quark fields with appropriate quantum numbers for the
baryons. 
For instance, 
the representation for baryons is given by a direct product of three
quarks~\cite{coji}
\begin{eqnarray}
\psi &\sim&
q\otimes q\otimes q \sim [ (\thalf,0)\oplus(0,\thalf) ]^3
\nonumber \\
\label{psiqqq}
&=& 5[(\thalf,0)\oplus(0,\thalf)] \oplus 3 [(1,\thalf)\oplus(\thalf,1)]
\oplus [(\hal{3},0)\oplus(0,\hal{3})]\, .
\end{eqnarray}
The multiplets $(\half,0)\oplus(0,\half)$ and
$(\hal{3},0)\oplus(0,\hal{3})$ have purely isospin $\half$ and
$\hal{3}$,
respectively, while $(1,\half)\oplus(\half,1)$ contains both $I=\half$
and $I=\hal{3}$.

It is natural to choose
the $(\half,0)\oplus (0,\half)$ representation
for the nucleon.
This multiplet transforms
in the same way as the quark field:
\begin{equation}
i \left[ Q_{R}^a , \psi_{r} \right] = - i t^a \psi_{r} \, \;\;\;\;\;
i \left[ Q_{L}^a , \psi_{l} \right] = - i t^a \psi_{l} \, .
\end{equation}
Here we assume to assign $(\half,0)$  to $\psi_r$.
However, it is also
possible to assign $(0,\half)$ to $\psi_r$. We shall discuss this
point in the next section.

For $\Delta$, which has $I=\hal{3}$, there are two candidates for the
representation of the chiral group (\ref{psiqqq}):
$[(1,\half)\oplus(\half,1)]$
and
$[(\hal{3},0)\oplus(0,\hal{3})]$. Here we
take the former multiplet, since the $\Delta$ is known as a strong
resonance of $N \pi$ system, and $N \otimes \pi\sim
[(\half,0)\oplus (0,\half)] \otimes [(\half,\half)] $ does not contain
$(\hal{3},0)\oplus(0,\hal{3})$. The transformation rules for a
field which
belongs to $[(1,\half)\oplus(\half,1)]$ are a little bit
complicated~\cite{Jido4}:
\begin{eqnarray}
\  [Q_R^A,(\psi_r)_i^B] &=& i \epsilon^{ABC} (\psi_r)^C_i \, , \;\;\;\;\;
  [Q_R^A,(\psi_l)_i^B] = -  (t^A)_{ij} (\psi_l)^B_j \, , \\
\  [Q_L^A,(\psi_r)_i^B] &=& -  (t^A)_{ij} (\psi_r)^B_j \, , \;\;\;\;\;
  [Q_L^A,(\psi_l)_i^B] = i \epsilon^{ABC} (\psi_l)^C_i\, .
\end{eqnarray}
The field $(\psi_{r,l})_{i}^A$ has two indices
for the chiral group.
Here $A=1,2,3$ is for $I=1$
triplet, and $i=1,2$ for $I=\half$ doublet.
It is important to note that this field contains both $I=\hal{3}$ and
$I=\half$ components. Therefore $\Delta$ and a nucleon resonance
$N^*$ form a representation of the chiral group.

From now on in this and the next sections,
we do not consider the quark structure of baryons and we
specify baryons as linear representations of the chiral group
$SU(2)_R \otimes SU(2)_L$.  
For most part of discussions, we consider 
the simplest representation of $(\half,0) \oplus (0, \half)$ for the 
nucleon and $(1,\half) \oplus (\half, 1)$ for the $\Delta$ and $N^*$.   
In reality, however, other representations could mix with this 
simplest representation~\cite{Weinberg1,Gerstein}.  
Therefore, results derived in the followings hold to the extent 
that the mixing is not considered large.

\subsection{Representations for two baryons}

Let us consider two kinds of baryons,
$\psi_{1}$ and $\psi_{2}$.
Their right and left handed components are denoted as
\beq
\psi_{1r}\, , \; \;
\psi_{1l}\, , \; \;
\psi_{2r}\, , \; \;
\psi_{2l}\, . \; \;
\eeq
To make argument explicit, we consider the nucleon which belongs
to $(\thalf,0)\oplus(0,\thalf)$. The following argument can be applied
to other representations with minor modifications.

We assume that $\psi_{1}$ belongs to the same representation as
that of (\ref{Qpsirlcomm}) or (\ref{VAtranspsi}).
Namely,
\beq
i \left[ Q_{R}^a , \psi_{1r} \right] = - i t^a \psi_{1r} \, ,
\; \; \;
i \left[ Q_{L}^a , \psi_{1l} \right] = - i t^a \psi_{1l} \, .
\label{Qpsi1rl}
\eeq
Now, for $\psi_{2}$ we can assume the same relation
\beq
i \left[ Q_{R}^a , \psi_{2r} \right] = - i t^a \psi_{2r} \, ,
\; \; \;
i \left[ Q_{L}^a , \psi_{2l} \right] = - i t^a \psi_{2l} \, .
\label{Qpsi2rlnai}
\eeq
We call the set of assignments of
(\ref{Qpsi1rl}) and (\ref{Qpsi2rlnai})
{\it naive}~\cite{Nemoto,Jido2}.

From chiral symmetry alone, however, it is also possible to
introduce another assignment for the second fermion,
\beq
i \left[ Q_{L}^a , \psi_{2r} \right] = - i t^a \psi_{2r} \, ,
\; \; \;
i \left[ Q_{R}^a , \psi_{2l} \right] = - i t^a \psi_{2l} \, .
\label{Qpsi2rlmirr}
\eeq
Namely, the right handed fermion transforms under $SU(2)_{L}$,  and
the left handed fermion transforms under $SU(2)_{R}$.
A somewhat confusing point stems, as anticipated,
from the fact that we have denoted
the chiral group as $SU(2)_{R} \otimes  SU(2)_{L}$ for the
right and left handed fermions.
Rather, we would have denoted as $SU(2)_{A} \otimes  SU(2)_{B}$.
Since the assignment of the two $SU(2)$ groups is interchanged for
the second fermion, we call the assignment
(\ref{Qpsi1rl}) and (\ref{Qpsi2rlmirr})
{\it mirror}~\cite{Jido2}.

In the four component representation, we can rewrite the
transformation rules of the naive and mirror assignments in the $VA$
representation as ($i = 1, 2$)
\beq
{\rm Naive:} \; \; \;  & &
i \left[ Q_{V}^a , \psi_{i} \right] = - i t^a \psi_{i} \, ,
\; \; \;
i \left[ Q_{A}^a , \psi_{i} \right] =  - i t^a \gamma_{5} \psi_{i} \, ,
\label{QVAnaiv}  \\
{\rm Mirror:} \; \; \; & &
i \left[ Q_{V}^a , \psi_{i} \right] = - i t^a \psi_{i} \, ,
\; \; \;
i \left[ Q_{A}^a , \psi_{i} \right] =  -
i  \eta  t^a \gamma_{5} \psi_{i} \, ,
\label{QVAmirr}
\eeq
where $\eta = 1$ for $i =1$ and $\eta = -1$ for $i = 2$.
These relations imply that the axial charge of $\psi_{2}$ in
the mirror assignment differs from that of $\psi_{1}$ in sign.

In order to distinguish the naive and mirror assignments,
we may indicate the chiral
group representations of the right
and left handed fermions explicitly:
\beq
\label{nairep}
{\rm Naive:} \; &\;&
\psi_{1} \sim \psi_{1r}(\thalf,0) + \psi_{1l}(0,\thalf) \, , \; \; \;
\psi_{2} \sim \psi_{2r}(\thalf,0) + \psi_{2l}(0,\thalf) \, , \\
\label{mirrrep}
{\rm Mirror:} \; &\;&
\psi_{1} \sim \psi_{1r}(\thalf,0) + \psi_{1l}(0,\thalf) \, , \; \; \;
\psi_{2} \sim \psi_{2r}(0,\thalf) + \psi_{2l}(\thalf,0) \, .
\eeq

From Lorentz and chiral invariance, one would have considered combinations
\beq
\label{psivec}
\psi_{r}(\thalf,0) + \psi_{l}(\thalf,0) \; \; {\rm or} \; \;
\psi_{r}(0,\thalf) + \psi_{l}(0,\thalf) \, .
\eeq
However, this assignment does not allow a Yukawa type meson-baryon 
interaction term.  
The chiral field is introduced with the transformation rule as defined  
in (\ref{sigmapi}).  
If two baryons are defined as the former of (\ref{psivec}) and are 
transformed by $g_{R}$ alone, it is obvious that the 
interaction term 
$\bar \psi (\sigma + i \vec \tau \scdot \vec \pi \gamma_{5}) \psi$
is not chiral invariant.  
If $\psi_{1}$ is defined by the former and $\psi_{2}$ by the latter 
of (\ref{psivec}), a Lorentz and chiral invariant term seems to be 
possible:
\beq
\psi_{2r}^\dagger U \psi_{1l}
+
\psi_{2l}^\dagger U \psi_{1r}
+
\psi_{1l}^\dagger U^\dagger \psi_{2r}
+
\psi_{1r}^\dagger U^\dagger \psi_{2l} \, ,
\eeq
where $U = \sigma + i \vec \tau \scdot \vec \pi$.  
This term, however, is not invariant under the parity transformation, 
$\psi_{12, r} \leftrightarrow \psi_{12, l}$ and
$U \leftrightarrow U^\dagger$.  
In this way, we see that the naive and mirror assignments are 
the two possible chiral assignments for meson and baryon systems.  



The transformation rules (\ref{QVAmirr})  for two isodoublet nucleons
were originally introduced in the textbook by Lee already
thirty years ago~\cite{Lee2}. There he constructed a linear sigma model
besed on that assignment.  
However his model
produced an unphysical consequence $g_{\pi NN}=0$,
since not all relevant terms allowed by chiral symmetry
were considered in his Lagrangian.  Then DeTar and Kunihiro
established a linear sigma model by adding terms missing in the
Lee's lagrangian and resolved the probrem there. 
Their model explained a possible parity doubling of
massive nucleons~\cite{DeTar}.
The DeTar-Kunihiro model was used in a study of finite density
properties of nuclear matter by Hatsuda and Prakash\cite{Hatsuda}.
Beane considered a similar model to the DeTar-Kunihiro model
in a different context in ref.\cite{beane}, where the mirror type assignment
is applied to quark fields. Christos also studied parity doubling
of nucleon using a local baryon field composed quark
fields\cite{christos}.

\subsection{Mass and parity eigenstates}

Let us consider a simple lagrangian
for one iso-doublet nucleon,
\beq
L = \sum_{f = p,n}
\left( \bar \psi_{f} i \dslash \psi_{f}
- m \bar \psi_{f}\psi_{f} \right) \, ,
\label{Lkandm1}
\eeq
where the sum over nucleon isospin is taken.
While the kinetic term preserves the chirality,
the mass term does not (here we omit writing the flavor indices):
\beq
L_{mass} &=& - m \psi_{r}^\dagger \psi_{l} - m \psi_{l}^\dagger \psi_{r}
\nonumber \\
&\to&
- m \psi_{r}^\dagger g_{R}^\dagger g_{L} \psi_{l}
- m \psi_{l}^\dagger g_{L}^\dagger g_{R} \psi_{r} \; \; \;
({\rm not \; invariant})\, ,  \nonumber
\eeq
where $g_{R} \in SU(2)_{R}$ and $g_{L} \in SU(2)_{R}$
are finite group elements of the chiral
group transformations as defined in (\ref{psiQRtr}).


Now we turn to the case of two iso-doublet baryons and consider a simple
lagrangian:
\beq
L = L_{K} + L_{M} \, ,
\eeq
where the kinetic term is given simply by the sum
\beq
\bar \psi_{1} i \dslash \psi_{1} +  \bar \psi_{2} i \dslash \psi_{2}
\, .
\eeq
From the Lorentz invariance and baryon number conservation,
possible forms of $L_{M}$ are
\beq
L_{M} &=&
- m_{1}
\left(
\psi_{1r}^\dagger \psi_{1l} +  \psi_{1l}^\dagger \psi_{1r}
\right)
- m_{2}
\left(
\psi_{2r}^\dagger \psi_{2l} +  \psi_{2l}^\dagger \psi_{2r}
\right)
\nonumber \\
&  & - m_{0}
\left(
\psi_{1r}^\dagger \psi_{2l} +  \psi_{2l}^\dagger \psi_{1r} +
\psi_{2r}^\dagger \psi_{1l} +  \psi_{1l}^\dagger \psi_{2r}
\right) \nonumber \\
&=&
- m_{1} \bar \psi_{1} \psi_{1}
- m_{2} \bar \psi_{2} \psi_{2}
- m_{0} (\bar \psi_{1} \psi_{2} + \bar \psi_{2} \psi_{1} )\, .
\label{LM2nai}
\eeq
For the naive baryons,  all the terms break
chiral symmetry.
Therefore, if we impose the theory to be explicitly chiral invariant,
the baryons must be massless.

For the mirror assignment, however, the third term of
(\ref{LM2nai}) preserves chiral
symmetry;
for instance,
\beq
\psi_{1r}^\dagger \psi_{2l} \to
(\psi_{1r}^\dagger g_{R}^\dagger) (g_{R} \psi_{2l})
= \psi_{1r}^\dagger \psi_{2l} \, . \; \; \;
{\rm (invariant)}
\eeq
Because of this,
the mirror baryons can acquire a (Dirac) mass keeping
chiral symmetry.
Since the mass term is off-diagonal in the particle space of $\psi_{1}$
and $\psi_{2}$, the
physical mass eigenstates are their linear combinations:
\beq
\psi_{+} = \frac{1}{\sqrt{2}}(\psi_{1} + \psi_{2}) \, , \; \; \;
\psi_{-} = \frac{1}{\sqrt{2}}(\psi_{1} - \psi_{2}) \, ,
\eeq
whose eigenvalues are $\pm m_{0}$.
The eigenvalue of $\psi_{-}$ appears negative, since $\psi_{-}$ carries
opposite parity of $\psi_{+}$.
Multiplying $\gamma_{5}$, the state $\gamma_{5} \psi_{-}$ carries the
same parity as $\psi_{+}$ and the eigenvalue turns to $+m_{0}$
(see Eq.~(\ref{massrev})).

Now we write the chiral transformation property in terms of
$\psi_{\pm}$:
\beq
\label{comm_mass}
i[ Q_{V}^a, \psi_{\pm} ] = -i t^a \psi_{\pm} \, ,
\; \; \;
i[ Q_{A}^a, \psi_{\pm} ] = -i t^a \gamma_{5} \psi_{\mp} \, .
\eeq
The second equation indicates that the mass eigenstates
$\psi_{\pm}$ form a large
representation of the chiral group.
It is, however, reducible, since the irreducible representations of
the chiral group can not diagonalize the mass matrix.
This is a unique feature of the mirror assignment 
and is a consequence that the 
eigenstates of chiral symmetry ($\psi_{1,2}$) differs from mass 
eigenstates, a situation analogous to quark mixing in the standard theory.

\section{Linear Sigma Models}

In this section we consider linear sigma models for
mesons and nucleons (denoted as $N$).
The meson fields are introduced as components of the representation
$(\half,\half)$
of the chiral group, which are subject to the transformation
rule:
\beq
\sigma + i \vec \tau \cdot \vec \pi
\to
g_{L} (\sigma + i \vec \tau \cdot \vec \pi) g_{R}^\dagger \, .
\label{sigmapi}
\eeq
The explicit symmetry breaking is not included in the following
discussion.

For later convenience, we note the following two properties.
First, the parity of the nucleon $N$ changes when multiplied by
$\gamma_{5}$.
This is because $\gamma_{5}$ anticommutes with
$\gamma_{0}$:
\beq
{\rm If}\; \; \;
P: N \to \gamma_{0} N \, , \; \; \;
{\rm then} \; \; \;
P: \gamma_{5}N \to - \gamma_{0} \gamma_{5} N\, .
\eeq
Second, the Dirac equation for $\gamma_{5}N$ changes the sign of $m$;
\beq
\label{massrev}
(i \dslash -m ) N = 0 \; \; \to \; \;
(i \dslash + m ) \gamma_{5} N = 0 \, .
\eeq

\subsection{The Gell-Mann Levy model}

Before going to the linear sigma models with two nucleons, we briefly
review some relevant aspects of 
the Gell-Mann Levy model with one nucleon belonging to
$(\thalf,0)\oplus(0,\thalf)$~\cite{Gell-Mann}.
The Lagrangian is given by
\begin{equation}
   {\cal L}_{GL} = i
\bar{N}  \dslash N
- g \bar{N}
(\sigma + i \gamma_5 \vec{\tau} \cdot \vec{\pi}) N
+ {L}_{\rm mes}  \, ,
\end{equation}
where $L_{\rm mes}$ denotes the Lagrangian for $\sigma$ and $\pi$. The
explicit form of $L_{\rm mes}$ is irrelevant in the following
discussion as long as it causes spontaneous
breaking of chiral symmetry.
The $\pi NN$ coupling is given simply by $g$ at
the tree level.
Once the spontaneous breakdown takes
place,
the mass of the nucleon is generated as
\begin{equation}
   m_N = g \sigma_0 \, , \label{massN}
\end{equation}
where
$\sigma_0 \equiv \bra 0 | \sigma | 0 \ket$
is the vacuum condensate of the sigma meson.
This equation is regarded as
the Goldberger-Treiman relation with $g_A=1$.

In the linear sigma model, it is allowed to add chirally
symmetric terms
containing derivatives and multi-mesons, once we abandon the
renormalizablity.
%
%
For instance,  one can add a term like
\begin{equation}
 \alpha \bar{N} [(\vec t \cdot \vec\pi \dslash \sigma - \sigma \vec t
 \dslash \vec\pi) \gamma_5 - \vec t \cdot
 (\vec \pi \otimes \dslash \vec\pi)] N \, .
\end{equation}
With this term $g_A$ is modified to be
$1 - \alpha \sigma_0^2$\cite{Lee2}.

Terms with multi-mesons give corrections to the nucleon
mass $m_N$ and the pion nucleon coupling $g$ as well as the axial 
charge in powers of $\sigma_0$. 
For instance, a chiral-invariant term
$\bar N
(\sigma^2+\pi^2)^n (\sigma + i \gamma_5 \vec\tau\cdot \vec\pi)N$ 
gives
contributions of order $\sigma_0^n$ and $n \sigma_0^{n-1}$ to $m_{N}$
and $g$, respectively.  
This should be contrasted with the non-linear sigma model which provides 
contributions in inverse powers of $\sigma_0$. 
Therefore, in the linear sigma model, the restoration of chiral symmetry 
is taken smoothly as a function of $\sigma_0 \to 0$, where correction 
terms disappear.  
From the experimental value $g_{A}^{exp} = 1.25$, the correction 
to $g_{A}$
from the higher derivative terms is of order of several tens percent.  

\subsection{Naive model}

First we consider the naive assignment.
It is not difficult to verify that up to forth order in
dimension of mass,
the general lagrangian can be written as~\cite{Jido2}:
\beq
L_{\rm{naive}}
& = &
\bar{N_1} i \dslash N_1
- g_{1} \bar{N_1}
(\sigma + i \gamma_5 \vec{\tau} \cdot \vec{\pi}) N_1
\nonumber \\
&+& \bar{N_2} i \dslash N_2
- g_{2} \bar{N_2}
(\sigma + i \gamma_5 \vec{\tau} \cdot \vec{\pi}) N_2
\nonumber \\
&-& \ g_{12} \{
\bar{N_1}
(\gamma_5 \sigma + i \vec{\tau} \cdot \vec{\pi}) N_2
-
\bar{N_2}
(\gamma_5 \sigma + i \vec{\tau} \cdot \vec{\pi}) N_1
\}
+ {L}_{\rm mes}  \, ,
   \label{ordsu2lag}
\eeq
where $g_{1}$, $g_{2}$ and $g_{12}$ are free parameters.
The terms of $g_{1}$ and $g_{2}$ are ordinary chiral invariant
coupling terms of the linear sigma model.
The term of $g_{12}$ is the mixing of $N_{1}$ and $N_{2}$.
Since the two nucleons have opposite parities, $\gamma_{5}$ appears
in the coupling with $\sigma$, while it does not in the coupling with
$\pi$.

Chiral symmetry breaks down spontaneously when the sigma meson
acquires a finite vacuum expectation value,
$\sigma_{0} \equiv \bra 0 |\sigma |0 \ket$.
This generates masses of the nucleons as in (\ref{massN}).
From (\ref{ordsu2lag}),
the mass can be expressed by a $2 \otimes 2$ matrix in the space
of $N_{1}$ and $N_{2}$.
Due to the mixing of different parity states, the diagonalization
has to be done carefully.
For this we introduce the new fields
\beq
(N^\prime_{1}, N^\prime_{2}) \equiv (N_{1}, \gamma_{5} N_{2}) \, ,
\eeq
in terms of which the original lagrangian can be written as:
\beq
{L}_{\rm{naive}}
& = &
\bar N^\prime_1 i \dslash N^\prime_1
- g_{1}
\bar N^\prime_1
(\sigma + i \gamma_5 \vec{\tau} \cdot \vec{\pi})
N^\prime_1
\nonumber \\
&+& \bar{N_2}^\prime i \dslash N_2^\prime
- g_{2}
\bar N^\prime_2 (\sigma + i \gamma_5 \vec{\tau} \cdot \vec{\pi})
N^\prime_2
\nonumber \\
&-& \ g_{12} \{
\bar N^\prime_1 ( \sigma + i \gamma_5 \vec{\tau} \cdot \vec{\pi})
N^\prime_2
+
\bar N^\prime_2 ( \sigma + i \gamma_5 \vec{\tau} \cdot \vec{\pi})
N^\prime_1
\}
+ \cdots \, .
\eeq
We note here the sign changes in the $g_{2}$ term and the second
term of $g_{12}$ terms.
The interaction terms, including the mass term,
are now expressed in the matrix form as
\beq
(\bar N^\prime_{1}, \bar N^\prime_{2} ) \sigma_{0} U_{5}
\left(
\begin{array}{c c }
    g_{1} & g_{12} \\
    g_{12} & - g_{2}
\end{array}
\right)
\left(
\begin{array}{c}
    N^\prime_{1} \\
    N^\prime_{2}
\end{array}
\right)\, ,
\eeq
where we have introduced
$U_{5}
= (\sigma + i \gamma_{5} \vec \tau \cdot \vec \pi )/\sigma_{0}$,
when
$\sigma_{0} \neq 0$.
This term can be diagonalized as
\beq
(\bar N^\prime_{+}, \bar N^\prime_{-} ) \sigma_{0} U_{5}
\left(
\begin{array}{c c }
    g_{+} & 0 \\
    0 & - g_{-}
\end{array}
\right)
\left(
\begin{array}{c}
    N^\prime_{+} \\
    N^\prime_{-}
\end{array}
\right)\, ,
\eeq
where
\beq
g_{\pm} = \frac{1}{2}
\left(
\sqrt{ (g_{1} + g_{2})^2 + 4g_{12}^2 }
\pm (g_{1} - g_{2} )
\right) \, ,
\eeq
and
\beq
    \left(
    \begin{array}{c}
	N_{+}^\prime  \\
	N_{-}^\prime
    \end{array}
    \right)
    =
    \left(
    \begin{array}{cc}
	\cos \theta      & \sin \theta  \\
	- \sin \theta & \cos \theta
    \end{array}
    \right)
    \left(
    \begin{array}{c}
	N_{1}^\prime  \\
	N_{2}^\prime
    \end{array}
    \right)  \, . \label{eigenpsi_nai}
\eeq
The mixing angle $\theta$ is defined by
\beq
\label{mixangle}
\tan 2\theta = \frac{2g_{12}}{g_{1} + g_{2}} \, .
\eeq

Turning back to the original fields, $N_{1,2}$, we find the physical
eigenfunctions:
\beq
    \left(
    \begin{array}{c}
	N_{+}  \\
	N_{-}
    \end{array}
    \right)
    =
    \left(
    \begin{array}{cc}
	\cos \theta      & \gamma_{5} \sin \theta  \\
	- \gamma_{5} \sin \theta &  \cos \theta
    \end{array}
    \right)
    \left(
    \begin{array}{c}
	N_{1}  \\
	N_{2}
    \end{array}
    \right)  \, , \label{eigenN_nai}
\eeq
whose masses are given by
\beq
m_{\pm} = \sigma_{0} g_{\pm} =
\frac{\sigma_{0}}{2}
\left(
\sqrt{ (g_{1} + g_{2})^2 + 4g_{12}^2 }
\pm (g_{1} - g_{2} )
\right) \, .
\label{massnaive}
\eeq

What is interesting here is that after the diagonalization, the two
nucleons decouple completely.
Explicitly, the lagrangian can be written as
\beq
{ L}_{\rm{naive}}
& = &
\bar{N_+} i \dslash N_+
- g_{+} \bar{N_+}
(\sigma + i \gamma_5 \vec{\tau} \cdot \vec{\pi}) N_+
\nonumber \\
& + &
\bar{N_-} i \dslash N_-
- g_{-} \bar{N_-}
(\sigma + i \gamma_5 \vec{\tau} \cdot \vec{\pi}) N_-
+ {L}_{\rm mes}  \, .
   \label{ldiagnai}
\eeq
Therefore, chiral symmetry puts no constraint on the relation
between $N_{+}$ and $N_{-}$.
The role of chiral symmetry is just the mass generation due to its
spontaneous breaking.
As shown in Fig.~\ref{mxangmass} (b),
when chiral symmetry is restored and
$\sigma_{0} \rightarrow 0$,
both $N_+$ and $N_-$ become massless and degenerate.
However, the degeneracy is trivial and has no relevance to chiral
symmetry between $N_{+}$ and $N_{-}$.
The decoupling of $N_{+}$ and $N_{-}$ implies that the
off-diagonal Yukawa coupling $g_{\pi N_{+}N_{-}}$ vanishes in the soft
pion limit.
This is valid up to the order we considered and if
derivative coupling terms are neglected.
The same result was also obtained in the QCD sum rule~\cite{Jido1}.

\subsection{Mirror model}

In the mirror assignment of chiral symmetry, the chiral invariant
lagrangian up to the order of (mass)$^4$ is given by
\begin{eqnarray}
{L}_{\rm{mirror}} & = &
\bar{N_1} i \dslash N_1
- g_{1} \bar{N_{1}}
(\sigma + i \gamma_5 \vec{\tau} \cdot \vec{\pi}) N_{1}
\nonumber \\
&+&
\bar{N_2} i \dslash N_2
- g_{2} \bar{N_{2}}
(\sigma - i \gamma_5 \vec{\tau} \cdot \vec{\pi}) N_{2}
\nonumber \\
&-& m_{0}( \bar{N_1} \gamma_{5} N_2 - \bar{N_2} \gamma_{5} N_1  )
+ {L}_{\rm mes} \ .
    \label{mirsu2lag}
\end{eqnarray}
In the $g_{2}$ term, the sign of the pion field is opposite
to that of the $g_{1}$ term, which ensures the chiral invariance of
the term in the mirror assignment.
This lagrangian was first formulated in the present form by DeTar and
Kunihiro~\footnote{There is a slight difference in the definition in
coupling constants. }.

When chiral symmetry is spontaneously broken, the mass matrix of the
lagrangian (\ref{mirsu2lag}) is given by
\beq
M =
\left(
\begin{array}{c c}
    g_{1}\sigma_{0} & m_{0}\gamma_{5} \\
    - m_{0}\gamma_{5} & g_{2} \sigma_{0}
\end{array}
\right) \, .
\label{mmatmirr}
\eeq
This can be diagonalized by a linear combination similar to
(\ref{eigenN_nai}) with the mixing angle:
\beq
\label{anglemirror}
\tan 2 \theta = \frac{2m_{0}}{\sigma_{0}(g_{1} + g_{2})} \, .
\eeq
The corresponding masses of the two states are given by
\beq
\label{massmirror}
m_{\pm} =
\half
\left(
\sqrt{ (g_{1} + g_{2})^2 \sigma_{0}^2 + 4m_{0}^2 }
\pm (g_{1} - g_{2} )\sigma_{0}
\right) \, .
\eeq
An interesting feature is that
the two nucleons become degenerate with a finite mass
$m_0$ when chiral symmetry is restored.

The mirror model contains
richer physics than the naive model, since the diagonalization
survives couplings between $N_{+}$ and $N_{-}$.
It is instructive to see first the axial charges which are then
related to the one pion Yukawa couplings through the Goldberger-Treiman
relation.
The axial charges are extracted from the commutation relations:
\beq
[Q_{A}^{a}, N_{+}] & = &  \hal{\tau^{a}} \gamma_{5}
(\cos 2 \theta \, N_{+}
- \sin 2 \theta \, \gamma_{5} N_{-})\, , \\
\ [Q_{A}^{a}, N_{-}] & = & \hal{\tau^{a}} \gamma_{5}
(- \sin 2 \theta \, \gamma_{5} N_{+} -
  \cos 2 \theta \,  N_{-}) \, ,
\label{mirq5com}
\eeq
giving a matrix expression
\beq
\label{gAmirror}
g_{A} = \left(
\begin{array}{c c }
    \cos 2 \theta & - \sin 2\theta\gamma_{5} \\
    - \sin 2\theta \gamma_{5} & - \cos 2 \theta
\end{array}
\right) \, .
\eeq

We would like to make several comments.  
\begin{enumerate}
    \item 
    As anticipated, the axial charge of the negative
    parity nucleon is just the minus of that of the positive parity
    nucleon.  
    This feature holds at any mixing angle $\theta$.  
    
    \item The absolute value of the diagonal component is, however, 
    less than unity.
    This may be increased by including higher derivative 
    terms.  
    
    \item
    The present result obtained from the linear sigma model can be 
    generalized by a group theoretical method, where the nucleon 
    axial charge
    $g_{A} \sim \bra N | Q_{A}^a | N\ket$
    is computed from the commutation relation
    $[Q_{A}^+, Q_{A}^-] = - Q_{V}$~\cite{Weinberg1,Gerstein}.   
    Assuming that the intermediate states of 
    \beq
    \bra N | Q_{A}^+  Q_{A}^- | N\ket =
    \sum_{B} \bra N | Q_{A}^+ | B\ket \bra B | Q_{A}^- | N\ket \, .
    \eeq
    are saturated by one baryon states, the method reproduces the 
    result obtained in the linear sigma model.  
\end{enumerate}

Now we extract the one pion Yukawa couplings from the lagrangian
written in terms of the physical basis $N_{+}$ and $N_{-}$.
The results are
\beq
\label{gpi_pp}
g_{\pi N_{+}N_{+}}
&=& g_{1} \cos ^2 \theta + g_{2} \sin ^2 \theta
= \frac{m_{+}}{\sigma_{0}} g_{A}^{++} \, , \\
\label{gpi_mm}
g_{\pi N_{-}N_{-}}
&=& - (g_{1} \sin ^2 \theta + g_{2} \cos ^2 \theta)
= \frac{m_{-}}{\sigma_{0}} g_{A}^{--} \, , \\
g_{\pi N_{+}N_{-}}
&=& \frac{g_{2} - g_{1}}{2} \sin 2 \theta
= \frac{m_{-} - m_{+}}{2\sigma_{0}} g_{A}^{+-} \, .
\eeq
The second equalities of these equations are the generalized
Goldberger-Treiman relations.

Finally we perform a simple parameter fitting~\cite{DeTar}.
In doing so, we assume to identify
$N_{+} \to N(939) \equiv N$ and $N_{-} \to N(1535) \equiv N^*$.
In order to fix the four parameters,
we use the inputs $m_{+} = 939, \; m_{-} = 1535$ MeV,
$\sigma_{0} = f_{\pi} = 93$ MeV and
$g_{\pi N_{+} N_{-}} \sim 0.7$.
The last equation is deduced from the partial decay width of
$N(1535)$~\cite{PDG} (although large
uncertainties for the width have been reported~\cite{ManSal,VrDtLe}):
\beq
\Gamma_{N^* \to \pi N} \sim 75 \; {\rm MeV} \, ,
\eeq
and the formula
\beq
\Gamma_{N^* \to \pi N}
= \frac{3q(E_{N} + M)}{4\pi M^*} g_{\pi NN^*}^2 \, ,
\eeq
where $q$ is the relative momentum
in the final state.
We find
\beq
\sigma_{0} &=& 93 \, {\rm MeV}\, , \; \; \;
m_{0} = 270 \, {\rm MeV}\, , \; \; \; \nonumber \\
\label{paramfit}
g_{1} &=& 9.8 \, , \; \; \; g_{2} = 16\, .
\eeq
From these parameters, we find the mixing angle
\beq
\theta = 6.3 \; \; {\rm degree} \, ,
\eeq
giving the diagonal value of the axial charges
\beq
\label{gApgAm}
g_{A}^{++} = - g_{A}^{--} = 0.98 \, .
\eeq

The different sign of the $g_{A}$'s as indicated in (\ref{gApgAm}) leads to 
different contributions in the calculation of the anomalous triangle 
diagram, when the loops of $N$ and $N^*$ are considered.   
The result may contradict if the anomaly matching is required as
an effective theory of QCD.  
However, it is argued that 
the anomaly should be accounted by quark loops rather
than baryon loops and  that the effective
lagrangian emerges when quark loops are integrated out.
Therefore, although not shown in our present lagrangians, the
Wess-Zumino-Witten term should be added
in order to account for the correct anomaly.  
In the mirror model, this interpretation is rather welcome,
since baryon loop contributions of $N$ and $N^*$ cancel due to the 
opposite signs of the $g_{A}$'s and the entire
anomaly is attributed to the quark loops.  In this regard,
the naive model is rather in contradiction as far as the anomaly 
matching to QCD is concerned, since baryon loops contribute 
constructively with the same sign of the $g_{A}$'s.

\subsection{Toward chiral symmetry restoration}

The change in chiral symmetry is dictated by the order parameter 
$\sigma_{0}$, which is identified with the pion decay constant
$f_{\pi}$.  
It decreases as temperature
or density is increased.
The naive and mirror models exhibit different behaviors of 
physical quantities such as masses and coupling constants as 
chiral symmetry is restored.  
The behavior of the naive case is rather trivial, and therefore, here 
we discuss mostly the mirror case.

\begin{figure}[tbp]
\centering
\epsfxsize = 12cm
\epsfbox{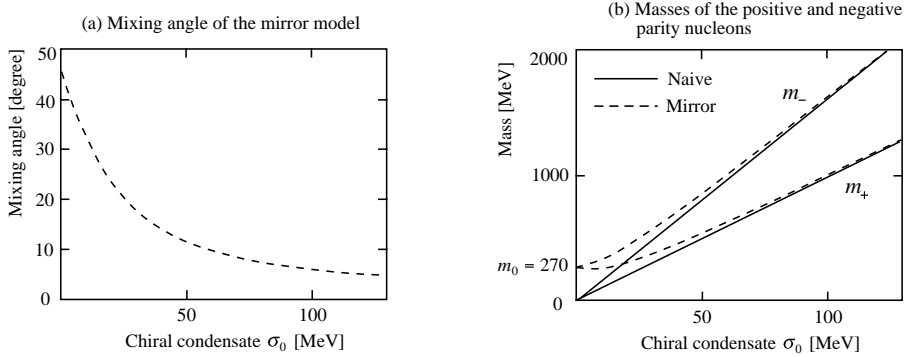}
\begin{minipage}{12cm}
   \caption{(a) Mixing angle of the mirror model, and
   (b) the masses of the positive and negative parity nucleons in the
   naive and mirror models, as functions of the chiral condensate
   $\sigma_{0}$.
      \label{mxangmass}}
\end{minipage}
\end{figure}
\begin{figure}[tbp]
\centering
\epsfxsize = 12cm
\epsfbox{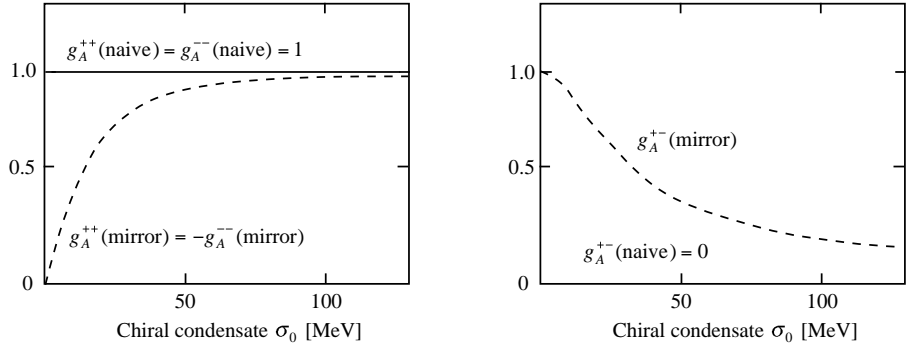}
\begin{minipage}{12cm}
   \caption{ Diagonal and off-diagonal axial charges in
   the naive and mirror models, as functions of the chiral condensate
   $\sigma_{0}$.
      \label{plotga}}
\end{minipage}
\end{figure}

\begin{itemize}
    \item Mixing angles and masses (Figs.~\ref{mxangmass})\\
    When chiral symmetry is restored ($\sigma_{0} = 0$), the only
    source
    of the mixing is the off-diagonal mass term (see (\ref{LM2nai}) or
    (\ref{mmatmirr})).
    Therefore, the two degenerate states $\psi_{1}$ and $\psi_{2}$
    mix with the equal weight ($\theta = \pi/4$) having an equal mass
    $m_{\pm} = m_{0}$.
    As the interaction is turned on and chiral symmetry starts to
    break spontaneously, the mixing angle decreases monotonically and
    reaches 6.3 degree in the real world ($\sigma_{0} = 93$ MeV).
    Also, as $f_{\pi}$ is increased 
    the masses increase and the degeneracy is resolved:
    $m_{+} = 939$ MeV and $m_{-} = 1535$ MeV when
    $\sigma_{0} = 93$ MeV.

    \item $g_{A}$ (Figs.~\ref{plotga})\\
    It is interesting to see the behavior of the diagonal
    $g_{A}^{++} = -g_{A}^{--}$ and off-diagonal
    $g_{A}^{+-}$ as functions of $\sigma_{0}$.
    In particular, $g_{A}^{+-}$ increases as chiral symmetry
    starts to restore.
    Since the axial charges are related to pion couplings, the
    increase of the off-diagonal couplings
    $g_{A}^{+-} \sim g_{\pi N_{+} N_{-}}$ was considered as
    a cause of the increase in the width of $N(1535)$ in
    medium~\cite{Kim,yamazaki,yorita}.
    More detailed calculation was reported in Ref.~\cite{Kim}.
\end{itemize}

\subsection{Higher representations}

Recently, Jido, Hatsuda and Kunihiro extended the simple mirror model
of fundamental representations to a model of higher dimensional
representations~\cite{Jido4}.
They considered a representation of $(1,\thalf) \oplus (\thalf,1)$.
The physics motivation of adopting it lies in the fact that the pion
couples strongly with the nucleon and delta.
A simple irreducible decomposition
\beq
\left(
(\thalf, 0) \oplus (0, \thalf)
\right)
\otimes
(\thalf, \thalf)
=
\left(
(1, \thalf) \oplus (\thalf, 1)
\right)
\oplus
\left(
(0, \thalf) \oplus (\thalf, 0)
\right) \, .
\label{irrdec}
\eeq
implies that the higher dimensional representation is realized by a
composite system of the pion and nucleon.
Incidentally, it is interesting to recognize that the second term of
(\ref{irrdec}) is the mirror representation of the original nucleon.
This is the basis of the five quark representation for the mirror
baryon as discussed in the next section.

Since the representation
$(1,\thalf) \oplus (\thalf,1)$
contains both isospin 1/2 and 3/2 components,
they have reached a {\it quartet scheme} in which $\Delta$, $N^*$ and
their parity partners form the chiral multiplet. Although their
argument can be applied for baryons with any spin $J$, to make
argument explicit,  here we take baryons with spin $J=\hal{3}$ as an
example. Then in the quartet scheme we have $\Delta_{+}(P_{33})$,
$\Delta_-(D_{33})$, $N^*_{+}(P_{13})$ and $N^*_-(D_{13})$.

In the mirror assignment, $\psi_{1r}$ and $\psi_{2l}$ belong to
$(1,\half)$, while $\psi_{1l}$ and $\psi_{2r}$ belong to
$(\half,1)$. Here the Lorentz index for the Rarita-Schwinger field
with $J=\hal{3}$ is omitted for simplicity. The transformation rules
under the chiral rotations are given by
\begin{eqnarray}
 \left[ Q_R^A, (\psi_{1r,2l})_i^B \right]
 &=&
i \epsilon^{ABC} (\psi_{1r,2l})^C_i\, ,
\nonumber \\
  \left[ Q_R^A, (\psi_{1l,2r})_i^B \right]
  &=& -  (t^A)_{ij} (\psi_{1l,2r})^B_j\, ,
  \nonumber\\
 \left[ Q_L^A,(\psi_{1r,2l})_i^B \right]
 &=& -  (t^A)_{ij} (\psi_{1r,2l})^B_j
\label{trhigher} \\
  \left[ Q_L^A,(\psi_{1l,2r})_i^B \right]
  &=& i \epsilon^{ABC} (\psi_{1l,2r})^C_i
  \, .  \nonumber
\end{eqnarray}
The reason that the naive assignment was not taken is that it gives an
unrealistic mass relation $m_{\Delta_\pm}=2m_{N^*_\mp}$.
The way to construct a chiral invariant Lagrangian is similar to what
we have explored here~\cite{Jido4}.
The interaction Lagrangian is
given by
\begin{eqnarray}
{\cal L}_{\rm int} &=& m_0 \left(\bar{\psi}^A_2 \gamma_5 \psi^A -
    \bar{\psi}^A_1 \gamma_5 \psi^A_2 \right) \nonumber \\
&+& a \bar\psi^A_1 \tau^B (\sigma - i \gamma_5 \vec\pi \cdot \vec\tau)
    \tau^A \psi_1^B +
  b \bar\psi^A_2 \tau^B (\sigma + i
\gamma_5 \vec\pi \cdot \vec\tau) \tau^A \psi_2^B\, .
\end{eqnarray}
The fields with the particle basis $ (\Delta_{1,2},N^*_{1,2})$ are
obtained by the following isospin decomposition:
\begin{equation}
  (\psi_{1,2})^A_i = \sum_M (T^A_\hal{3})_{iM}\Delta_{1,2}^M + \sum_m
  (T^A_\half)_{im} N^{*m}_{1,2} \, ,
\end{equation}
where the projection matrices are defined through the Clebsch-Gordan
coefficients:
\beq
(T^A_I)_{im} &=&
\sum_{\mu \nu}
(1 \mu \, \thalf \nu | I m) \epsilon_{\mu}^A \chi_{\nu}^i \, ,
\eeq
where $(\epsilon_{\mu}^A) = \vec \epsilon_{\mu}$ are polarization
vectors
\beq
\vec \epsilon_{+1} =
- \frac{1}{\sqrt{2}}
\left(
\begin{array}{c}
    1\\
    i\\
    0
\end{array}
\right) \, , \; \; \;
\vec \epsilon_{-1} =
\frac{1}{\sqrt{2}}
\left(
\begin{array}{c}
    1\\
    -i\\
    0
\end{array}
\right) \, , \; \; \;
\vec \epsilon_{0} =
\left(
\begin{array}{c}
    0\\
    0\\
    1
\end{array}
\right) \,
\eeq
and $(\chi_{\nu}^i) = \chi_{\nu}$ spinors
\beq
\chi_{1/2} =
\left(
\begin{array}{c}
    1\\
    0
\end{array}
\right) \, , \; \; \;
\chi_{-1/2} =
\left(
\begin{array}{c}
    0\\
    1
\end{array}
\right) \, .
\eeq

After spontaneous breakdown of chiral symmetry and
diagonalization of the mass matrices, the masses of the $\Delta$'s and
$N^*$'s are given by
\begin{eqnarray}
  m_{\Delta_\pm} &=& \sqrt{(a+b)^2 \sigma_0^2 + m_0^2} \mp \sigma_0
  (a-b) \, , \\
  m_{N^*_\pm} &=& \sqrt{\left({a+b \over 2}\right)^2 \sigma_0^2 +
  m_0^2} \pm {\sigma_0 \over 2}  (a-b)\, .
\end{eqnarray}
These equations show that the chiral symmetry breaking induces the
mass splitting between parity partners as well as between isospin states.
It is interesting to note that these equations lead to  parameter
free and nontrivial relations among baryon masses:
\beq
{\rm sgn} (m_{\Delta_{+}} - m_{\Delta_{-}})
&=& - {\rm sgn} (m_{N_{+}^*} - m_{N_{-}^*})\, , \nonumber \\
\half
(m_{\Delta_{-}} - m_{\Delta_{+}})
&=&
m_{N_{+}^*} - m_{N_{-}^*} \, , \label{massformula} \\
\half
(m_{\Delta_{+}} + m_{\Delta_{-}})
&\ge&
\half
(m_{N_{+}^*} + m_{N_{-}^*}) \, .\nonumber
\eeq

As mentioned before, the present argument holds for the resonances
with an arbitrary spin as far as we consider the $(1,\half)\oplus
(\half,1)$ multiplet.
Here we assign the quartet to the lowest
resonances in each spin parity among the observed baryons. However, for
$J=\half$ the nucleon is assumed to belong to $(\half,0)\oplus
(0,\half)$ with a possible parity partner, $N(1535)$ or
$N(1650)$ (or even their linear combinations).
Therefore, we investigate two quartet schemes for $J=\half$:
that including $N(1535)$ (case 1) or that including $N(1650)$ (case 2).
In table~\ref{tb:quartet}, the parameter free relations
(\ref{massformula}) are compared with those expected from
observed masses.

\begin{table}[tbp]
 \caption{Comparison between parameter free
 predictions of the  quartet scheme (QS)  and the data~\cite{Jido4}.
 Case 1 and case 2 in the $J=\hal{1}$
 sector stand for  the cases where $N^{*}_{-}$=$N(1535)$ and
$N^{*}_{-}$=$N(1650)$, respectively. The last two
  rows are model parameters $m_0,a$ and $b$
  determined from the experimental inputs.}
 \begin{center}
  \begin{tabular}{|c|c|c|c|c|c|}
  \hline
       &  QS & \multicolumn{2}{c|}{$J=\hal{1}$}
       &  $J=\hal{3}$  & $J=\hal{5}$  \\
       & & case 1 & case 2 & & \\
    \hline
    ${\rm sgn}\left( { m_{N_{+}^{*}} - m_{N_{-}^{*}} \over
                m_{\Delta_{+}}- m_{\Delta_{-}} } \right) $
       &  $-$ & $-$  & $-$   & $-$   &  $-$  \\
    \hline
    ${ m_{N_{+}^{*}} - m_{N_{-}^{*}} \over
                m_{\Delta_{+}}- m_{\Delta_{-}} } $
       & $-\hal{1}$   &  $-0.33$ &  $-0.72$  &  $-0.43$  &  $-0.2$  \\
    \hline
   ${ m_{N_{+}^{*}} + m_{N_{-}^{*}} \over
                m_{\Delta_{+}}+ m_{\Delta_{-}} } $
       & $ \le 1 $  &  0.84 & 0.88 & 1.1   &  0.87  \\
    \hline \hline
   \multicolumn{2}{|c|} {$m_0$ (MeV)} & 1380 & 1460 & 1540 & 1590 \\
   \hline
    \multicolumn{2}{|c|} {$(a, \ b)$}  & $(5.2, 6.6)$ & $(4.4, 6.1)$
   & $(1.2,-1.2)$ & $(5.8, 5.7)$ \\
 \hline
  \end{tabular}
 \end{center}
 \label{tb:quartet}
\end{table}

\begin{figure}[btp]
    \centering
    \footnotesize
   \epsfxsize=8.0cm
    \epsfbox{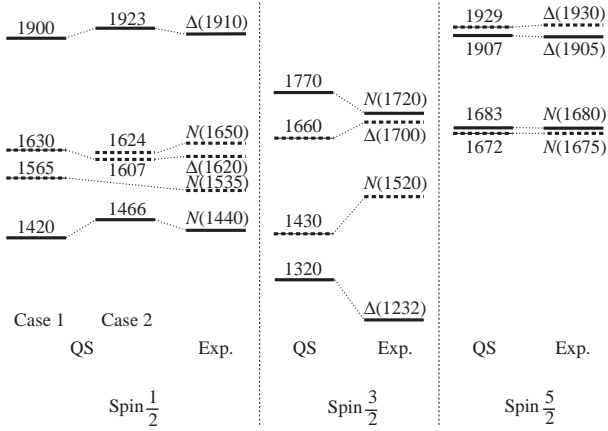}
    \caption{The quartet members with 
    $J=\hal{1},\hal{3},\hal{5}$~\cite{Jido4}. The
    right (left)  hand side for each spin
    is the observed (quartet scheme) masses. The solid (dashed) lines
denote
    the even (odd) parity baryons. The reproduced masses in our scheme
    agree with the experimental values within 10 \%.}
    \label{quartet}
\end{figure}

Encouraged by the agreement of the parameter free relations, we
determine three model parameters.
This is done  for each quartet of different spin.
We take $f_{\pi} = 93$ MeV, and then
four experimental masses are used for
least square fitting.
The resulting parameters are presented in table~\ref{tb:quartet}.
In Fig.\ref{quartet} masses of the
quartet members are compared with experiments~\cite{Jido4}.
It is shown that the experimental masses are fitted
within 10\%.

In the $J=\hal{3}$ the fitted parameters
$a=-b=1.2$ lead approximately
to interesting selection rules for the $\pi$-decay
of the the quartet baryons.
If $a=-b$, the interaction Lagrangian of
$\pi$ and $\psi_\pm$ is written as
\begin{equation}
   {\cal L}_{1\pi} = (\bar{\psi}^A_+,\bar{\psi}^A_-)
   \left(
     \begin{array}{cc}
        0 & -a \\
        a & 0
     \end{array}
   \right) \tau^B (i \vec\pi \cdot \vec\tau) \tau^A
   \left(
     \begin{array}{c}
        \psi^B_+ \\
        \psi^B_-
     \end{array}
   \right)
\end{equation}
where $\psi_+ = {1\over \sqrt{2}}(\psi_1 + \gamma_5 \psi_2)$ and
$\psi_- = {1\over \sqrt{2}}(\gamma_5 \psi_1 -  \psi_2)$. This
interaction Lagrangian implies that the parity nonchanging couplings,
such as $\pi \Delta_{\pm} \Delta_{\pm}$, $\pi \Delta_{\pm} N^*_\pm$
and $\pi N^*_\pm N^*_\pm$, are forbidden at the tree level. This is
qualitatively consistent with the decay patterns of the observed
resonances. These suppressions, in fact,  have been seen also in the
constituent quark model\cite{koniuk} and the collective string
model\cite{iachello}, although the considered symmetry and dynamics
are quite different from the present model. Therefore it is interesting
to compare the physical reason and the relations among the models to
understand the role of chiral symmetry in the baryon resonances.

\section{QCD Spectral Function of Positive and Negative Parity Baryons}

We have seen that the low energy effective theory of hadrons is strongly
constrained by chiral symmetry assignment of hadrons.  As  QCD is
the origin of the effective theory, the assignment must be consistent
with QCD.  We face two crucial questions: (1) which assignment is
consistent with  QCD for chiral symmetry of baryons? and (2)
how can we confirm the assignment for actual baryon resonances?

In order to answer these questions, we consider the following two steps.
First we construct local operators  (interpolating fields) for hadrons
in terms of quark fields as components of a suitable chiral group
representation.
Then, as the second step, we consider chiral properties of hadron
states which are coupled to the local operators.
Since chiral symmetry is spontaneously
broken and the QCD vacuum is not chiral invariant, the second step is
nontrivial.
We consider two-point correlation functions of baryons
and how the chiral symmetry of the local operators is connected to
that of the baryonic states.

\subsection{Interpolating fields}

Under the $SU(2)_{R} \otimes SU(2)_{L}$ chiral transform, the quark
operator $q(x)$ behaves as the fundamental representation,
$(\half,0)\oplus(0,\half)$ (see Eq.~(\ref{VAtranspsi})):
\begin{equation}
    [Q_A^a , q(x)] = - \gamma_5 t^a\,q(x) \, ,
\end{equation}
where $t^a = \tau^a /2$ are the generators of $SU(2)$.
The quark field is an $SU(2)$ isospin doublet and is denoted as
$$  q(x)= \textstyle{\left(\matrix{u(x)\cr d(x)}\right)} \, .$$
Here
the Dirac indices of quark fields and gamma matrices are suppressed.

To start with, let us see how interpolating fields for mesons are
constructed as desired chiral group representations.
The pion is special as the Nambu-Goldstone boson of
spontaneous breaking of chiral symmetry.  The PCAC (in the strong
sense) conjectures a widely accepted relation,
$$
\partial_{\mu} A^{a\mu}(x)
= m_{\pi}^{2} f_{\pi} \phi_{\pi}^{a}(x) \, .
$$
Then the pion corresponds to a local quark bilinear operator,
$$
\phi_{\pi}^{a}(x) \sim
\partial_{\mu}[\qbar(x) \g{\mu} \gamma_5 \tau^{a} q(x)]_{\rm{CS}}
\sim 2m_{q}[\qbar(x) i\gamma_{5} \tau^{a} q(x)]_{\rm{CS}} \, .
$$
Here $[\qbar\ldots q]_{\rm{CS}}$ means that $(\qbar\ldots q)$ is made
color-singlet by summing over the color indices.
It is easy to see that this operator is transformed as the $a$-th
component of a chiral $O(4)$ vector.  The axial transformation gives
\begin{equation}
    \null[Q_A^a, [\qbar i\gamma_{5}
    \tau^b q]_{\rm{CS}}] = i \delta^{ab}
    [\qbar q]_{\rm{CS}}
    \quad:\qquad
    \null[Q_A^a, [\qbar q]_{\rm{CS}}] = -i [\qbar i\gamma_{5}
    \tau^a q]_{\rm{CS}}
    \label{eq:SPcomm}
\end{equation}
Here $[\qbar q]_{\rm{CS}}$ is the fourth component of the $O(4)$
vector,  which corresponds to a scalar isoscalar meson, i.e. $\sigma$.
Thus the set of operators,
\begin{equation}
    \left( [\qbar(x) i\gamma_{5} \tau^{a} q(x)]_{\rm{CS}}\, ,
    \; [\qbar(x) q(x)]_{\rm{CS}} \right)
    \sim (\pi^a \, , \; \sigma)\, ,
    \label{eq:SPdef}
\end{equation}
belong to a linear representation $(\half, \half)$ of the chiral
$SU(2)_{R} \otimes SU(2)_{L}$ group.
The mass splitting of $\pi$ and
$\sigma$ is therefore a manifestation of the spontaneous breaking of
chiral symmetry.

Similarly, one can introduce vector mesons as two
independent representations:  $(1,0)$
and $(0,1)$.  They are represented by
$[\qbar_{l}(x) \g{\mu}\tau^{a}  q_{l}(x)]_{\rm{CS}} $ and
$[\qbar_{r}(x) \g{\mu}\tau^{a}  q_{r}(x)]_{\rm{CS}} $,
where the right
and left projections are defined by (\ref{psirandl}).
Each of these representation is not an eigenstate of the parity, but
the $r\pm l$ combinations give the vector and axial-vector
operators of parity eigenstates.
Again in the chiral limit, they must be degenerate, while the physical
mesons corresponding to these currents are $\rho$ and $a_{1}$, which
are split by about 400 MeV.

Now we consider the nucleon operator, which is supposed to consist of
three quark operators.
A choice of such an operator is~\cite{Ioffe}
\begin{equation}
    B^{\alpha} (x) \equiv \left[ (q^T(x) C \gamma_{5}\,q(x))_{I=0} q(x)^{\alpha}
    \right]_{\rm{CS}}
    \label{eq:B}
\end{equation}
where $\alpha$ is the Dirac index of the nucleon and $C$ is the charge
conjugation gamma matrix, $C= i \gamma^2\gamma^0$ (standard representation).
This is a local operator with the
appropriate quantum number for the nucleon, $S=\half$ and $I=\half$.
The first two quarks in (\ref{eq:B}) form a scalar di-quark with $I=0$,
which commutes with $Q_A^{a}$.  Thus the
nucleon operator itself behaves just like a quark field under chiral
transform,
\begin{equation}
    [Q_A^a, B(x)] = - \gamma_{5} t^a B(x)
    \label{eq:QB}
\end{equation}
This equation indicates that $B$ belongs to the linear representation,
$(\half,0)\oplus(0,\half)$, with the axial charge $g_A=1$.

There is an alternative choice for the nucleon operator,
\begin{equation}
    \Bt^{\alpha} (x) \equiv
    \left[ (q^T(x) C q(x))_{I=0} \{\gamma_{5}\,q(x)\}^{\alpha}
    \right]_{\rm{CS}} \, .
    \label{eq:BT}
\end{equation}
It is easy to see that $\Bt$ also satisfies eq.(\ref{eq:QB}), and
therefore has $g_A=1$.
In the following arguments we need not distinguish $\Bt$ from $B$.
Indeed, most general local operator (without derivative) for the proton
is given explicitly by~\cite{Espriu}
\begin{equation}
    B_p(x) = \epsilon^{abc} \left[ (u^{a\ T} C d^b)\,\gamma_{5} u^c
     + t (u^{a\ T} C\gamma_{5}\,d^b) \,u^c \right] \, ,
     \label{bpgeneral}
\end{equation}
where $a$, $b$, and $c$ stand for the color of the quarks, and $t$
is a parameter which controls the mixing of the two independent
operators.

In the previous sections,
we have pointed out the possibility that an excited
nucleon is a mirror partner of the ground state.
The signature of the mirror assignment is the
negative axial charge $g_{A}$.
We have seen that the three-quark baryon operators,
(\ref{eq:B}) and (\ref{eq:BT}), give the naive nucleon.
It turns out that the mirror baryon consists of more than three quarks
if it is constructed as a local operator without derivatives.

We consider the color-singlet scalar/pseudoscalar bilinear operators,
given by (\ref{eq:SPdef}), which are transformed as (\ref{eq:SPcomm}).
Then, the mirror baryon can be represented by a local 5-quark operator,
\begin{equation}
    B^{*\alpha} = [\qbar q ]_{\rm{CS}} B^{\alpha}
    + [\qbar i\gamma_{5} \tau^b q]_{\rm{CS}}
    (i \gamma_{5} \tau^b B)^{\alpha} \, ,
    \label{eq:Bmirror}
\end{equation}
where $B$ is defined by either (\ref{eq:B}) or (\ref{eq:BT}).
Using eqs.(\ref{eq:QB}) and (\ref{eq:SPcomm}),
one can check
\begin{equation}
    [Q_A^a, B^{*}] = +\gamma_5 t^a B^* \, ,
\end{equation}
and thus $B^*$ has a negative axial charge $g_A = -1$.
We conclude that $B^{*}$ represents a mirror nucleon.

\subsection{Correlation functions}

If chiral symmetry were realized in the Wigner mode,
i.e. not spontaneously broken, in QCD, then the operators with
different chirality would not mix and would generate distinct hadron
states.  Unfortunately (or otherwise) chiral symmetry is broken.

For example, consider matrix elements,
\begin{equation}
    \langle 0|[\qbar i\gamma_{5} \tau^b q]_{\rm{CS}} |\pi^{a}\rangle,
    \; \; \;
    {\rm and}
    \; \; \;
   \langle 0|[\qbar q]_{\rm{CS}} |\sigma\rangle \, .
\label{eq:SPmatt}
\end{equation}
Using eq.(\ref{eq:SPcomm}), we find
$$ \langle 0|[\qbar i\gamma_{5} \tau^a q]_{\rm{CS}} |\pi^{a}\rangle =
\langle 0|i [Q_{A}^{a} ,[\qbar q]_{\rm{CS}}] |\pi^{a}\rangle \, .
$$
In the Wigner mode, $Q_{A}^{a}|0\rangle =0$ and $(\pi^{a}, \sigma)$
form a $O(4)$ vector.  Therefore, we find
$$ \langle 0|[\qbar i\gamma_{5} \tau^a q]_{\rm{CS}} |\pi^{a}\rangle =
- \langle 0|i  [\qbar q]_{\rm{CS}} Q_{A}^{a} |\pi^{a}\rangle =
+ \langle 0| [\qbar q]_{\rm{CS}} |\sigma\rangle \, .
$$
Thus, the matrix elements, (\ref{eq:SPmatt}), are related to each
other.
This is, however, not the case if the symmetry is spontaneously
broken. $Q_{5}^{a}$ does not annihilate the vacuum and the pion and
sigma states are not simply related by the axial transformation.
Instead, if we apply the soft pion theorem~\cite{DoGoHo92},
we obtain in the limit $p_{\pi}\to 0$,
$$ \langle 0|[\qbar i\gamma_{5} \tau^a q]_{\rm{CS}} |\pi^{a}\rangle =
-  {i\over f_{\pi}}
\langle 0| [Q_{A}^{a} ,[\qbar i\gamma_{5} \tau^a q]_{\rm{CS}} ] |0\rangle
=  {1\over f_{\pi}} \langle 0| [\qbar q]_{\rm{CS}} |0\rangle \, .
$$
Thus the pion matrix element is related to the $\qbar q$ condensate in the
vacuum. This relation leads to the Gell-Mann-Oakes-Renner relation.

In the baryon sector, we expect that the symmetry breaking in the
vacuum mixes different chiral representations.
Furthermore, it should be noted that the local operator
(\ref{eq:B}) or (\ref{eq:BT})
does not create only the ground state nucleon but also couples to excited
states.  It is therefore necessary to single out information of the
ground state in order to study the properties of the nucleon.
A standard technique to achieve this separation is
to consider a two-point correlator of the above operators~\cite{JidoSR},
\begin{equation}
    T(p) \equiv i\int d^4x\, e^{ipx} \theta (x^0) \langle 0| B(x) \bar
    B(0) |0\rangle = \gamma^0 A(\sqrt{s}) + B(\sqrt{s}) \, ,
    \label{eq:2point}
\end{equation}
where we take the reference frame $p^0=\sqrt{s}$, and $\vec p=0$.
It is easy to see that the spectral representation of this
correlation function contains
contributions of both the positive and negative parity baryons.
That is,
\begin{equation}
    T(p) = {1\over \pi} \int {1+\gamma^0\over \sqrt{s} -m-i\epsilon}\,
    \rho^{(+)}(m) \, dm -
    {1\over \pi} \int {1-\gamma^0\over \sqrt{s} -m-i\epsilon}\,
    \rho^{(-)}(m) \, dm
\end{equation}
where the first (second) term represents the contribution of the
positive (negative) parity baryon spectrum.
Using the dispersion relation, we find
\begin{equation}
    \rho^{(\pm)}(m) = {1\over 2} \,\Im [ A(m) \pm B(m)]
\end{equation}
We immediately see that $\rho^{(\pm)}(m) $ are equal if $B(m)= 0$,
while the observed baryon spectrum shows that they are not.

Eq.(\ref{eq:2point}) is often employed in constructing QCD sum
rules~\cite{JidoSR}.
The correlation function can be evaluated in the deep Euclid momentum
region, i.e., $-(p^0)^{2} \to \infty$, resulting an expression in terms of
vacuum expectation values of various local operators, such as
the quark condensate and gluon condensate.
Then using the dispersion relation, we derive a relation of the spectral
function $\rho$ to the QCD vacuum properties.

Applying the QCD sum rule to (\ref{eq:2point}) we find that chiral
symmetry breaking is responsible for the splitting of positive and
negative  parity
baryons.
It turns out that the QCD part of $\Im B(m)$  contains the local
operators which break chiral symmetry, such as $\langle\qbar q
\rangle$,  $\langle \qbar \sigma_{\mu \nu} G^{\mu \nu} q \rangle$.
Thus one sees directly that chiral symmetry breaking makes $B(m)$
nonzero and therefore produces the mass difference of the positive
and negative parity baryons.

A careful QCD Borel sum rule analysis with standard values of the
quark condensate and other QCD parameters confirms that the spectrum
of the positive and negative parity octet baryons,\cite{JidoSR} i.e.,
the nucleon, $\Lambda$, $\Sigma$ and $\Xi$, is fairly well reproduced.
It is also shown that the flavor singlet baryon shows a special
behavior so that the negative parity state has a lower mass than its
positive parity partner.  The lower state is assigned to the lightest
observed negative-parity strange-baryon $\Lambda(1405)$.

This result shows that the mass splittings of the positive and
negative parity baryons are attributed to chiral symmetry breaking.
Can we tell whether they are destined to be mirror baryons in chiral
symmetry restoration?  In Eq.(\ref{eq:2point}), we employ the naive
baryon operators, {\it i.e.} three-quark operators of
(\ref{bpgeneral}).
We have seen that
their correlation function gives excited baryon spectrum that agrees
with experiment reasonably.  However, as the chiral symmetry
breaking mixes chiral representations in general, we do not
immediately conclude that the excited baryons which are coupled to
(\ref{bpgeneral}) are ``naive''.  Further,
we may have more than one baryon resonances in a relatively small
energy region, as is for $1/2^{-} N^{*}$.  This unfortunate situation
prevents us from making a definite conclusion on the chiral properties
of the observed baryon resonances.

However, we can study the properties of the baryon operators by using
the correlation functions.  Two point correlation functions of the
mirror nucleon operator reveal interesting features on the
pion-nucleon couplings.  Especially, the off-diagonal coupling of the
pion to the positive parity and negative parity nucleons can be shown
to vanish for the naive baryons.  It can be seen in the soft-pion
limit of the two-point correlator with a background pion:
\begin{eqnarray}
    T_{\pi}^{a}(p) &=& \lim_{q\to 0} i \int d^{4}x \, e^{ip\cdot x}
     \langle 0| {\rm T} (B(x) B^{*}(0)) |\pi^{a}(q)\rangle
     \nonumber\\
     &=& i( A_{\pi}^{a}(p^{2}) p\cdot\gamma \gamma_{5} +
     B_{\pi}^{a}(p^{2})\gamma_{5} )
     \nonumber\\
     &=& -{i\over f_{\pi}} \int d^{4}x \, e^{ip\cdot x}
     \langle 0| [Q_{5}^{a}, {\rm T} (B(x) B^{*}(0))] |0 \rangle \, .
    \label{eq:pi-correlator}
\end{eqnarray}
The same correlator can be expressed in terms of the $\pi B B^{*}$
coupling constant by
\begin{equation}
    T_{\pi}^{a}(p)  \simeq g_{\pi B B^{*}} \lambda_{B}\lambda_{B^{*}}
    {m_{B}+m_{B^{*}}\over (p^{2}-m_{B}^{2})(p^{2}-m_{B^{*}}^{2})}
    ip\cdot\gamma \gamma_{5}\tau^{a} + \cdots
    \label{eq:pi-BB}
\end{equation}
for the negative parity baryon excitation $B^{*}$.
This shows that  $A(p^{2})$ in (\ref{eq:pi-correlator}) is
the term responsible for the off-diagonal $\pi B B^{*}$ coupling.
Substituting the 3-quark or 5-quark operators for $B$ and $B^{*}$
in  (\ref{eq:pi-correlator}), we find that for the naive $B^{*}$
the term $A$ vanishes identically in the soft-pion
limit~\cite{Jido1}, while
for the mirror $B^{*}$ it is related to the chiral-even term of the
$B B^{*}$ correlator in the vacuum.

This result is consistent with the linear sigma model described in
section 3.
Namely, in the naive case,
as the $\pi$-baryon coupling comes from the term that is
responsible for the baryon masses, the coupling is diagonal for the
mass eigenstate of baryons.

In conclusion, further studies of the baryonic correlation functions,
both in QCD sum rules and in lattice QCD calculations will be useful.
It is, however, not straightforward to answer the questions like
whether a particular nucleon excited state is a mirror of the ground
state (or the other excited state) baryon and  how much different chiral
symmetry representations are mixed in the excited states.

\section{Observation of Chiral Properties of Baryons}

\subsection{Introduction}

In this section, we study possible reactions to distinguish our two
models for nucleons,
which have been discussed in the previous sections.
For this we propose two meson productions of $\pi$ and $\eta$ at the
threshold region induced by either the pion or photon.

As discussed in the preceding sections, one of the differences
between the naive and mirror assignments is the relative
sign of the axial charges of the positive and negative
parity nucleons.
In the following discussions, we identify $N_{+} \sim N(939) \equiv N$
and $N_{-} \sim N(1535) \equiv N^*$.
This identification for the nucleons is no more than an assumption.
From experimental point of view, however, $N(1535)$ has an advantage
that it has a strong coupling to an $\eta$ meson, which
can be used as a signature for the resonance.

\begin{figure}[tbp]
\centering
\epsfxsize = 8cm
\epsfbox{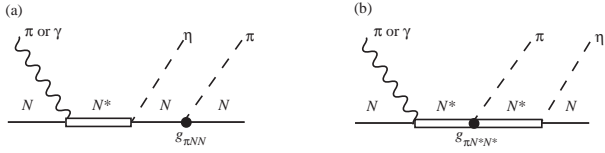}
\begin{minipage}{12cm}
   \caption{
   Two diagrams for $\pi$ and $\eta$ productions. The incident
   particle denoted by wavy line is either pion or photon.
   \label{twodiag}}
\end{minipage}
\end{figure}

In the reactions we propose,
the relative sign of the axial charges
may be observed by interference effects.
In fact, the axial charges are related to the pion Yukawa couplings
through the Goldberger-Treiman relations (\ref{gpi_pp}) and
(\ref{gpi_mm}).
Hence the relative sign of $g_{\pi NN}$ and $g_{\pi N^*N^*}$ follows
that of $g_{A}^{NN} (\equiv g_{A}^{++})$ and
$g_{A}^{N^*N^*} (\equiv g_{A}^{--})$.
The essential idea may be explained by using  two
diagrams shown in
Fig.~\ref{twodiag}.
The incident particle can be either the pion or photon.
Let us assume that the two diagrams dominate the
reaction close to the threshold.
The diagrams (1) and (2) in
Fig.~\ref{twodiag} have the coupling $\gpNN$ and $\gpRR$,
respectively. Therefore the two terms interfere either constructively
or destructively,  depending on the relative sign of the couplings.
In reality, due to kinematical conditions, other diagrams will
contribute, but we will find that there are some cross
sections in which interference effect can be seen clearly.


\subsection{Pion induced reaction:  $\pi + N \to \pi + \eta + N$}

First we investigate the pion induced process.
Here we consider $p\pi^- \rightarrow p\pi^-\eta$ process, since
there are two charged particles in the final state, which has an
advantage in the experimental setup. If the emitted proton and pion
are observed and their invariant mass is measured, it is not necessary
to observed the neutral eta meson. $p\pi^- \rightarrow n\pi^0\eta$ is
also possible, in principle, to observe the relative sign, but there
are more neutral particles. $p\pi^+ \rightarrow p\pi^+\eta$ cannot be
used, since $p\pi^+$ state has purely isospin $I=3/2$
and cannot couple with $N(1535)$ before emitting a pion. Therefore there
is no chance to enter $\gpRR$ in the process.

\begin{figure}[tbp]
    \vspace*{1cm}
    \centering
    \footnotesize
    \epsfxsize = 10cm
    \epsfbox{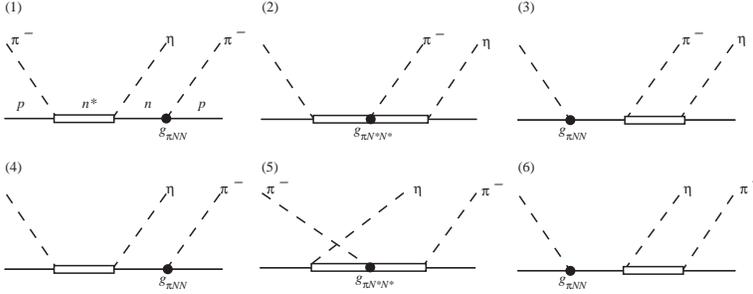}
    \begin{minipage}{12cm}
    \centering
    \caption{ \small
    Six $N^*$ dominant diagrams for $\pi^- p \rightarrow
    \pi^-\eta p$ process.
    \label{6diag}}
    \end{minipage}
\end{figure}

We assume $N^* = N(1535)$ dominance as in
single $\eta$ productions, where
it is known that the $\eta$ couples strongly between $N$ and  $N^*$,
while other couplings are negligibly small.
In this $N^*$ dominance, we can write altogether
six diagrams as shown
in Figs.~\ref{6diag}.

If we strictly follow the linear sigma model lagrangians in section 3,
the diagrams involving $\pi NN^*$ vanishes for
the naive assignment.  
However, if we consider small explicit breaking 
of chiral symmetry and/or higher order derivative terms, corrections 
yield the small coupling $g_{\pi NN^*} \sim 0.7$ which is 
extracted from the decay width
$\Gamma_{N^* \to \pi N} \sim 75$ MeV.  
is introduced in the naive 
Hence in this section we consider the difference of the naive and the
mirror model only in the relative sign of the $\pi NN$ and $\pi N^* N^*$
coupling constants.

In this spirit, we adopt the following interaction lagrangians for
actual computations of cross sections:
\beq
L_{\pi NN}
&=& \gpNN \bar N i \gamma_{5} \vec \tau \cdot \vec \pi N \, ,
\nonumber \\
L_{\eta NN^*}
&=& \geNR ( \bar N \eta N^*  + \bar N^* \eta N )  \, ,
\nonumber \\
L_{\pi NN^*}
&=& \gpNR ( \bar N \tau \cdot \pi  N^*
+ \bar N^* \tau \cdot \pi N )  \, ,
\nonumber \\
L_{\pi N^*N^*}
&=& \gpRR ( \bar N^* i \gamma_{5} \tau \cdot \pi  N^* )  \, .
\label{Lints}
\eeq
We use empirical values for
coupling constant; $\gpNN \sim 13$, $\gpNR \sim 0.7$
and $\geNR \sim 2$.
The unknown parameter is the $\gpRR$ coupling.
One can estimate it by using the theoretical value of the
axial charge $g_{A}^{N^* N^*}$ and the Goldberger-Treiman relation for
$N^*$.
When $g_{A}^{N^* N^*} = \pm 1$ for the naive and mirror assignments,
we find
$
\gpRR = g_{A}^{N^* N^*}\  m_{N^*}/{f_{\pi}} \sim \pm 17 .
$
On the other hand,  a recent work in the chiral unitary
approach gives $\gpRR \sim 5$\cite{nacher}.
Here, just for simplicity, we use the same absolute value as
$g_{\pi NN}$.
The coupling values
used in our computations are summarized in
Table~\ref{parameters}.

\begin{table}[tbp]
    \centering
    \caption{\label{parameters} \small Parameters used in our
calculation. }
    \vspace*{0.5cm}
    \begin{tabular}{ c c c c c c c }
    \hline
    $m_{N}$ & $m_{N^*}$ & $\Gamma_{N^*}$ & $\gpNN$
    & $\gpNR$ & $\geNR$ & $\gpRR$  \\
    \hline
    938 & 1535  & 140            & 13
    & 0.7     &  2.0    &  13 (naive) \\
    (MeV) & (MeV) & (MeV) &
    &         &         &  --13 (mirror) \\
    \hline
    \end{tabular}
\end{table}

\begin{figure}[tbp]
    \vspace*{1cm}
    \centering
    \footnotesize
    \epsfxsize = 11cm
    \epsfbox{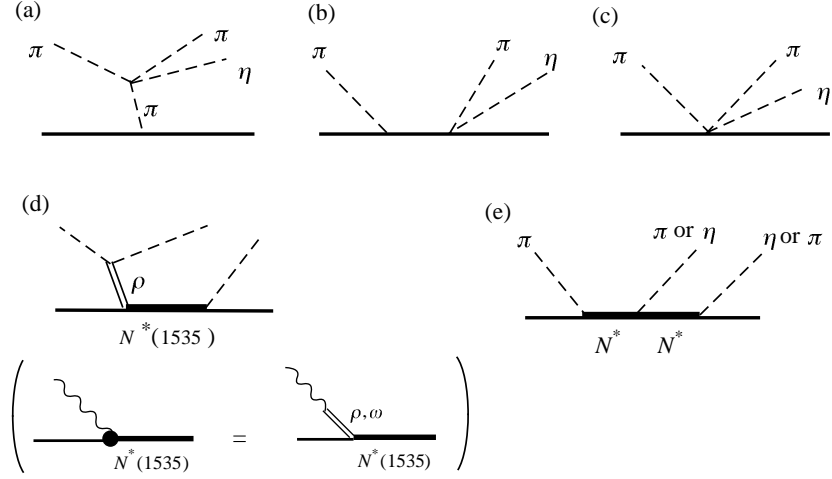}
    \begin{minipage}{12cm}
    \centering
    \caption{ \small
    Various contributions to $\pi N \to \eta \pi N$.
    \label{bkground}}
 \end{minipage}
\end{figure}

Besides the resonance dominant contributions as shown in
Fig.~\ref{6diag}, there are several other possible terms.
We ignore all of them from the following reasons.
\begin{itemize}
    \item  {\bf Background:}\\
    First we consider three diagrams as shown in Figs.~\ref{bkground}
    (a-c).
    All of them are not allowed due to G-parity.
    For (a), the vertex $\pi \pi \pi \eta$ is G-parity forbidden, since
    $G(\pi) = -1$ and $G(\eta) = 1$.
    For (b) and (c), to estimate the diagrams, we first consider
    the lowest order of the chiral expansion in the
    lagrangian~\cite{DoGoHo92}.
    For two-meson nucleon vertices in (b), the two mesons are
    correlated as vector mesons (such as $\rho$) which have G-parity
    plus.
    Therefore, the G-parity minus combination $\pi \eta$ is not
    allowed.
    Similarly, three meson vertices in (c) have axial vector
    correlation with negative G-parity,
    and hence the positive G-parity combination $\pi \pi \eta$ is
    not allowed.
    These selection rules are explicitly satisfied in chiral
    lagrangians.

    \item  {\bf $\rho$ meson:}\\
    We have computed the diagram in Fig.~\ref{bkground} (d) explicitly.
    The rho meson coupling to $N$ and $N^*$ is extracted from the helicity
    amplitudes $A_{1/2} \sim 0.08 \; GeV^{-1/2}$~\cite{PDG}
    using the vector meson dominance as shown in Fig.~\ref{bkground} (d).
    It turns out that the contribution to the cross section
    is negligibly small as compared
    to the resonance pole terms in Fig.~\ref{12diag}
    by about a factor $10^{-3}$.

    \item  {\bf Other resonance contributions:}\\
    Finally, one would expect contributions where two resonances appear
    in intermediate states as shown in Fig.~\ref{bkground} (e).
    Again we can ignore these diagrams safely,
    since there is no strong indication
    that baryon resonances couple to $N^*(1535)$ by emitting a
    pion~\cite{PDG}.
    In particular, the delta resonance which could be excited strongly
    by the incident pion couples only weakly to $N^*(1535)$, since the
    observed branching ratio of $N^*(1535) \to \Delta \pi$ is less than
    10 \% \cite{PDG}. The contamination from resonances of isospin
    $I=3/2$ may be estimated by $p\pi^+ \rightarrow p \pi^+ \eta$.
\end{itemize}

\begin{figure}[tbp]
    \vspace*{1cm}
    \centering
    \footnotesize
    \epsfxsize = 7cm
    \epsfbox{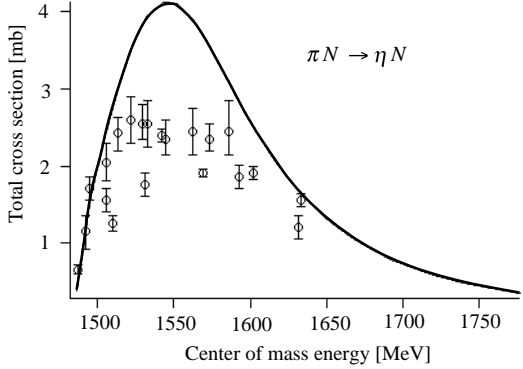}
    \begin{minipage}{12cm}
    \centering
    \caption{ \small
    Total cross section of $\pi N \to \eta N$.
    Data are taken from~\cite{Batinic}.
    \label{pinetan}}
 \end{minipage}
\end{figure}

Before discussing the process we are interested in, we mention
that our present approach does not include initial state interaction
which affects and reduce absolute values of cross sections
by factor of a few times.
We show in Fig.~\ref{pinetan} the total cross section of
$\pi + N \to \eta + N$, where theoretical cross sections computed
using the present lagrangian and parameters as compared with
experiments.
Theoretical cross section is indeed smaller than experimental one
by a factor nearly two at the peak position.

\begin{figure}[tbp]
    \vspace*{1cm}
    \centering
    \footnotesize
    \epsfxsize = 6cm
    \epsfbox{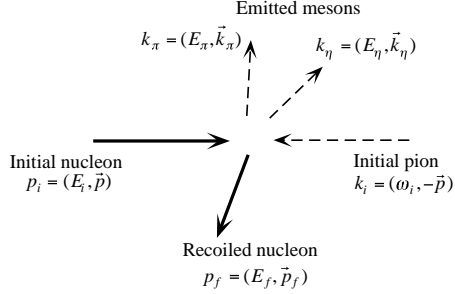}
    \begin{minipage}{12cm}
    \centering
    \caption{ \small
    Definition of momentum variables in the center of mass (CM) frame.
    \label{momenta}}
    \end{minipage}
\end{figure}

Now we compute the $S$-matrices.
Reduction formula allows us to write
\beq
S_{fi} &=&
_{out}\bra p_{f}, k_{\pi}, k_{\eta} | k_{i}, p_{i}\ket_{in}
\nonumber \\
&=&
disc.  + (iZ_{\pi}^{-1/2})^2 (iZ_{\eta}^{-1/2})
\int d^4x d^4y d^4z \,
e^{ik_{\pi}x + ik_{\eta}y - ik_{i} z} \nonumber \\
& & (\Box_{x} + m_{\pi}^2)  (\Box_{y} + m_{\eta}^2)
(\Box_{z} + m_{\pi}^2)
_{out}\bra p_{f}|T(\pi^i(x) \eta(y) \pi^j(z)) | p_{i}\ket_{in} \, .
\label{smatrix}
\eeq
The momentum variables are defined as shown
in Fig.~\ref{momenta}.
In the perturbation theory, the $T$-matrix element can be computed as
\beq
& & _{out}\bra p_{f}|T(\pi^i(x) \eta(y) \pi^j(z
)) | p_{i}\ket_{in}
\nonumber \\
& & \; = \;
i^3 \bra p_{f}|T ( \pi^i(x) \eta(y) \pi^j(z)
\int d^4x_{1} d^4x_{2} d^4x_{3}
L_{\eta NN^*}(x_{1}) L_{\pi NN^*}(x_{2}) \nonumber \\
& & \hspace*{2cm}
\otimes (L_{\pi NN}(x_{3}) + L_{\pi N^*N^*}(x_{3})) )
| p_{i}\ket_{in} \, .
\label{ptbn}
\eeq
After performing the Wick contraction in (\ref{ptbn}) which is then
inserted in (\ref{smatrix}), we find an ordinary expression for
amplitudes in momentum space.
For instance the amplitudes for Fig.~\ref{12diag}
(1) and (2) are given by
\begin{eqnarray}
    {T}{(1)} &=& \bar{u}(p_{f}) { (i\sqrt{2}g_{\pi NN}i\gamma_{5})i
    (i g_{\eta NN^{*}}) i( i \sqrt{2} g_{\pi NN^{*}})\over
    (\pslash_{f} + \kslash_{\pi} - m_{N})( \pslash_{i}+\kslash_{i} -
    m_{N^{*}} + \hal{i} \Gamma )}u(p_{i}) \, , \\
    {T}{(2)} &=& \bar{u}(p_{f}) {(i g_{\eta NN^{*}} ) i(i
    \sqrt{2} g_{\pi N^{*}N^{*}} i \gamma_{5})i (i \sqrt{2} g_{\pi
    NN^{*}}) \over (\pslash_{f} + \kslash_{\eta} - m_{N^{*}} + \hal{i}
    \Gamma) (\pslash_{i} + \kslash_{i} - m_{N^{*}} +\hal{i} \Gamma)
    }u(p_{i}) \, ,
\end{eqnarray}
where $u$'s are the Dirac spinors for the nucleon
and
$\Gamma = \Gamma_{N^*(1535) \to \pi N}
+ \Gamma_{N^*(1535) \to \eta N} = 140$ MeV is the total width of $N(1535)$.

Using the $T$-matrix,
${\cal T}_{fi} = T(1) + \cdots T(6)$, we calculate differential
cross section as
\beq
d\sigma &=& \frac{2m_{N}}
{4 \sqrt{(p_{i} \cdot k_{i})^2 - m_{N}^2 m_{\pi}^2}} \frac{1}{2}
\sum_{spin} |{\cal T}_{fi} |^2 d \Phi \, ,
\label{crosssection}
\eeq
where the phase space of the three body final state is given by
\begin{equation}
    d\Phi =
    (2\pi)^{4}  \delta(p_{i}+k_{i}-p_{f}-k_{\pi}-k_{\eta})
    {d^{3}k_{\pi} \over (2\pi)^{3} 2 E_{\pi}}
    {d^{3}k_{\eta} \over (2\pi)^{3} 2 E_{\eta}}
    {m_{N} d^{3}p_{f} \over (2\pi)^{3} E_{f}} \, .
\end{equation}
Here the convention for the normalization is
\begin{eqnarray}
    \bar{u}^{(\alpha)}(p) u^{(\beta)}(p)  =  \delta^{\alpha\beta}
     \, ,  \; \; \;
     \bracket{p}{p^{\prime}} = {E \over m} (2\pi)^{3} \delta^{3}
     (\vec{p}-\vec{p}^{\prime}) \, .
\end{eqnarray}
In the center of mass frame, the phase space integral measure reduces
to\cite{PDG}
\begin{equation}
    d\Phi = \frac{m_{N}}{4(2\pi)^{5}} dE_{\pi} dE_{f}
    d\alpha d(\cos\beta) d\gamma \, .
\end{equation}
Here the orientations of the momenta of the emitted particles are
specified by the three Euler angles $\alpha$, $\beta$ and $\gamma$,
since, in the center of mass frame, the three momenta,
$\vec{p}_{f}$ $\vec{k}_{\pi}$ and $\vec{k}_{\eta}$, lie in a common
plane.
Then the relative angles between either two of $\vec{p}_{f}$,
$\vec{k}_{\pi}$ and $\vec{k}_{\eta}$ can be determined, if the
energies of the proton and the pion in the final state, $E_{f}$,
$E_{\pi}$ are fixed.

We have computed the integral over the three body phase space in
the Monte Carlo method.
The number of configurations is taken more than 30,000, depending
on the kinds of cross section.
The total cross section is computed in a schematic way as
\begin{eqnarray}
    \sigma  =  {\rm (K.F.)} \int |{\cal T(\xi)}|^{2}  d\Phi
     \to   {\rm (K.F.)}{1 \over N} \sum_{i=1}^{N}
    |{\cal T}(\xi_{i})|^{2} V\, ,
\end{eqnarray}
where $\xi=(E_{f}, E_{\pi}, \alpha, \cos\beta, \gamma)$.
The volume of the phase space is given by the integral
\begin{equation}
    \label{volint}
    V = {m_{N} \over 4 (2 \pi)^{5}} 4 \pi^{2} \int_{E_{min}}^{E_{max}}
    {2 \sqrt{(E^{*2}_{\eta}- m_{\eta}^{2})(E^{*2}_{\pi}-m_{\pi}^{2})}
    \over E_{cm}} dE_{\pi} \, ,
\end{equation}
where $E^{*}_{\eta}$ and $E^{*}_{\pi}$ are the energies of the emitted
$\eta$
and $\pi$ in the rest frame of the emitted nucleon and eta.
They are
\begin{eqnarray}
    E^{*}_{\eta} & = & { E_{cm}^{2}-m_{N}^{2} + m_{\eta}^{2} +
    m_{\pi}^{2} - 2 E_{cm} E_{\pi} \over 2\sqrt{E_{cm}^{2} +
    m_{\pi}^{2} - 2E_{cm} E_{\pi}}} \, , \\
    E^{*}_{\pi} & = & {E_{cm} E_{\pi} - m_{\pi}^{2} \over
    \sqrt{E_{cm}^{2} + m_{\pi}^{2} - 2 E_{cm} E_{\pi}}} \, .
\end{eqnarray}
The lower and upper bounds of the integral (\ref{volint})
are given by
\begin{eqnarray}
    E_{min} & = & m_{\pi} \, , \\
    E_{max} & = & { E_{cm}^{2} + m_{N}^{2} - (m_{N} + m_{\eta})^{2}
    \over 2 E_{cm} }\, .
\end{eqnarray}

To compute differential cross section $d\sigma(\zeta)$ where
$\zeta$ is a representative of the variable we need,
for instance, the angle of the emitted pion in the center of mass
frame  and
the momentum of the emitted pion in the laboratory frame.
We insert a delta function of finite range $\Delta \zeta$ in the
integrand of the total cross section.
\begin{eqnarray}
    d\sigma (\zeta)& = &{\rm (K.F.)} \int |{\cal T}(\xi)|^2
    \delta(\zeta^\prime (\xi) - \zeta) d \Phi \nonumber \\
    & \to &  {\rm (K.F.)}{1 \over N} \sum_{i=1}^{N}
    |{\cal T}(\xi_{i})|^{2} { 1 \over \sqrt{\pi}  \Delta\zeta}
    e^{-{(\zeta^\prime(\xi_i)-\zeta)^2 \over \Delta\zeta^2}}
    V\, .
\end{eqnarray}
Note that the phase space is represented in the center of mass frame.
The $\zeta^\prime$ stands for the translation of the variables $\xi$.
For example, $\zeta^\prime$ is a boost transformation from the CM frame to
the laboratory frame to calculate the differential cross section in the
laboratory frame.

\begin{figure}[tbp]
    \vspace*{1cm}
    \centering
    \footnotesize
    \epsfxsize = 12cm
    \epsfbox{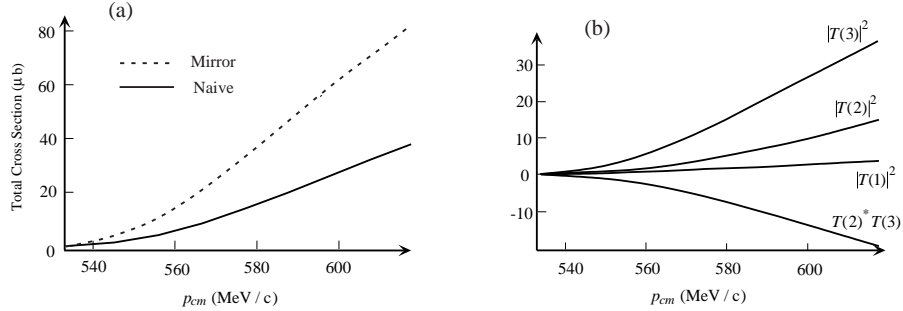}
    \begin{minipage}{12cm}
    \centering
    \caption{ \small
        (a) Total cross sections of
    $\pi^- p \to \eta \pi^- p$ for the naive and mirror models as
    functions of the initial pion momentum
    $P_{c.m.}$ in the center of mass frame.
    (b) Separate contributions from various terms of Fig.~\ref{12diag}.
    The amplitude of the cross term $T(2)^{*} T(3)$ corresponds to the
    naive assignment.  It changes the sign for the mirror assignment.
    \label{crtot}}
    \end{minipage}
\end{figure}

The total cross sections is shown in Fig.~\ref{crtot} (a)
as functions of
the energy of the initial pion for the naive and mirror cases.
The difference between the two is due to the sign
of the $\gpRR$ coupling.
The cross sections increase, as the initial pion energy and
correspondingly the phase space of the final three body state
increase.
For $P_{cm} \gsim P_{cm}^{\rm threshold} + 50$ MeV/c
($P_{cm}^{\rm threshold} = 528$ MeV/c), the
cross sections reach more than ten micro barn, which
will be experimentally accessible.
In the whole energy region as shown in Fig.~\ref{crtot}
the cross section is larger in the mirror model,
which is about twice as that in the naive model.

Among various terms shown in Fig.~\ref{12diag}, major contributions
are from the diagrams (2) and (3).
In Fig.~\ref{crtot} (b), we show relative strengths of
$|T(2)|^2$, $|T(3)|^2$ and their interference $2 T(2)^{*} T(3)$ in the
naive assignment.
The difference between the naive and mirror assignments in the total
cross section is from the sign difference of this term.
The third major contribution is from $T(1)$ which is also shown
there.
Other terms are negligible.

In the pion induced reaction,
the term $T(3)$ gives a large contribution
to the cross section as compared
with $T(1)$ and $T(2)$, although we expected naively that
the terms $T(1)$ and $T(2)$ were the dominant contributions when
considering their energy denominators. This can be explained by
the p-wave nature of the $\pi NN$ and $\pi N^* N^*$ couplings.
In the diagrams 1 and 2, a p-wave pion is emitted with
a small momentum, which suppresses the amplitudes.
In the diagram (3), a p-wave pion is absorbed with
a large momentum, which enhances the amplitude.

%
%
%

\begin{figure}[tbp]
    \vspace*{1cm}
    \centering
    \footnotesize
    \epsfxsize = 13cm
    \epsfbox{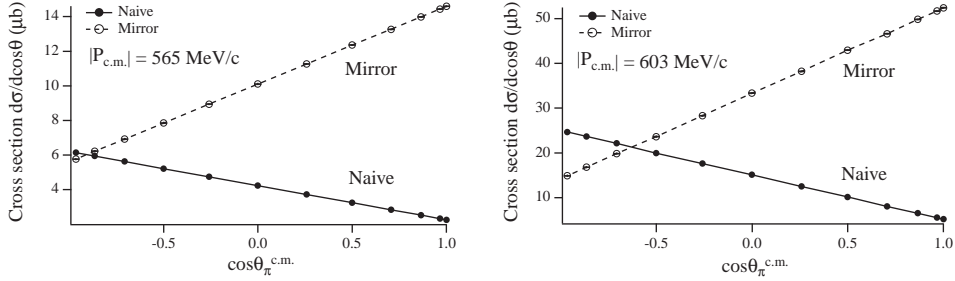}
    \begin{minipage}{12cm}
    \centering
    \caption{ \small
    Angular distributions of the
    $\pi^-$ in the final state. \label{ang}}
    \end{minipage}
\end{figure}

\begin{figure}[tbp]
    \vspace*{1cm}
    \centering
    \footnotesize
    \epsfxsize = 13cm
    \epsfbox{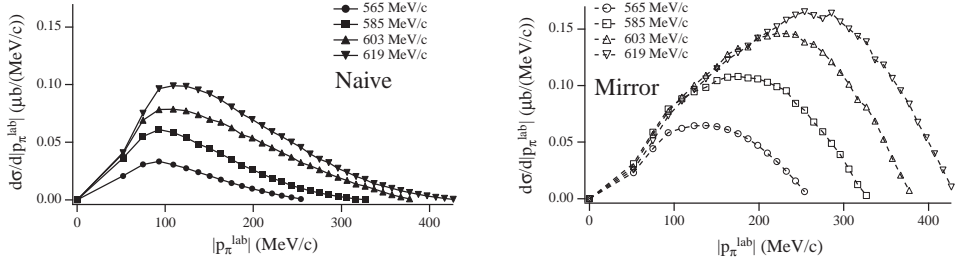}
    \begin{minipage}{14cm}
    \centering
    \caption{ \small
    Momentum distributions of the
    $\pi^-$ in the final state. \label{piene}}
    \end{minipage}
\end{figure}

In the present calculation, we have not considered
the initial and final state interactions between the pion and
nucleon.
The inclusion of realistic interactions
might reduce absolute values of cross sections by an overall factor
of several times, as we have discussed the single eta production
$p \pi^- \rightarrow n \eta$ in Fig.~\ref{pinetan}.
It is important to consider such distortion
effect when we discuss absolute values of the cross sections.
However, relative behaviors of such as momentum and angular
distributions are not affected by this effect.
These are now discussed as follows.

\begin{itemize}
    \item  {\bf Angular distributions:}
    In Fig.~\ref{ang}, we show angular distributions of the final
    state  pion in the center of mass frame, which
    shows  clear difference between the two models.
    This comes
    from the p-wave nature of the $\pi N^*N^*$ coupling in the
    diagram (2) of Figs.~\ref{12diag}.
    As it is mentioned before, the main contributions are given by the
    diagram (2) and (3), and the
    difference in the sign of $\gpRR$ appears in the
    sign of the interference term of $T(2)$ and $T(3)$.
    Due to the structure of the Yukawa vertex, the emitted pion
    in the diagram (2) is in
    p-state, which gives linear dependence on
    $\cos \theta_{\pi}^{c.m.}$.
    On the other hand, the emitted pion in
    the diagram (3) is in s-state, which
    has no angular distributions.
    Therefore the cross term of $T(2)$ and $T(3)$ behaves
    linearly in $\cos \theta_{\pi}^{c.m.}$, and
    the apparently different behaviors in the
    $\cos \theta_{\pi}^{c.m.}$
    dependence is due to the difference in the sign
    of the coupling $\gpRR$.
    This angular dependence would be one of the cleanest observables to
    distinguish the naive or mirror assignments.

    \item  {\bf Momentum distributions:}
    Another example which is useful
    is the momentum (energy) distribution of one of the final state
    particles.
    We plot the momentum distribution
    of the emitted pion in the laboratory frame in
    Fig.~\ref{piene} for several incident energies.
    What differs in
    the two chiral assignments is the position of the
    peak in the cross sections.
    In the naive case, it does not depend on
    the incident energy very much,
    while in the mirror case, it shifts to higher
    momentum region as the incident energy is increased.
\end{itemize}

\subsection{Photon induced reaction:  $\gamma + N \to \pi + \eta + N$}

In general, photoproductions contain richer
physics than pion induced processes, since the photon carries spin
one and has isoscalar and isovector components.
Our investigation is similar to what was explored for the two pion 
productions~\cite{takaki,nacherdpp}, but the process considered here 
is a little bit simpler since the eta meson is an isoscalar particle 
and exclusively couples to $N(1535)$.  
Under reasonable assumptions, we can follow very closely to the method 
for the pion induced reactions.
In the following we discuss several
items relevant to the photoproduction.
Detailes will be reported elsewhere~\cite{johprep}.

\begin{figure}[tbp]
    \vspace*{1cm}
    \centering
    \footnotesize
    \epsfxsize = 4cm
    \epsfbox{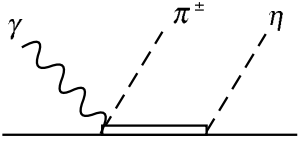}
    \begin{minipage}{14cm}
	\centering
	\caption{ \small
	A diagram of the Kroll-Ruderman type for a charged pion production.
	\label{kroll}}
    \end{minipage}
\end{figure}

First, as in the case of single pion photoproduction,
we expect that the Kroll-Ruderman term as shown in Fig.~\ref{kroll}
plays a dominant role for charged pion production.
However, this term is suppressed for a neutral pion production.
For this reason, in the following discussions we consider the neutral pion
production: $\gamma p \rightarrow \pi^0 \eta p$.

\begin{figure}[tbp]
    \vspace*{1cm}
    \centering
    \footnotesize
    \epsfxsize = 12cm
    \epsfbox{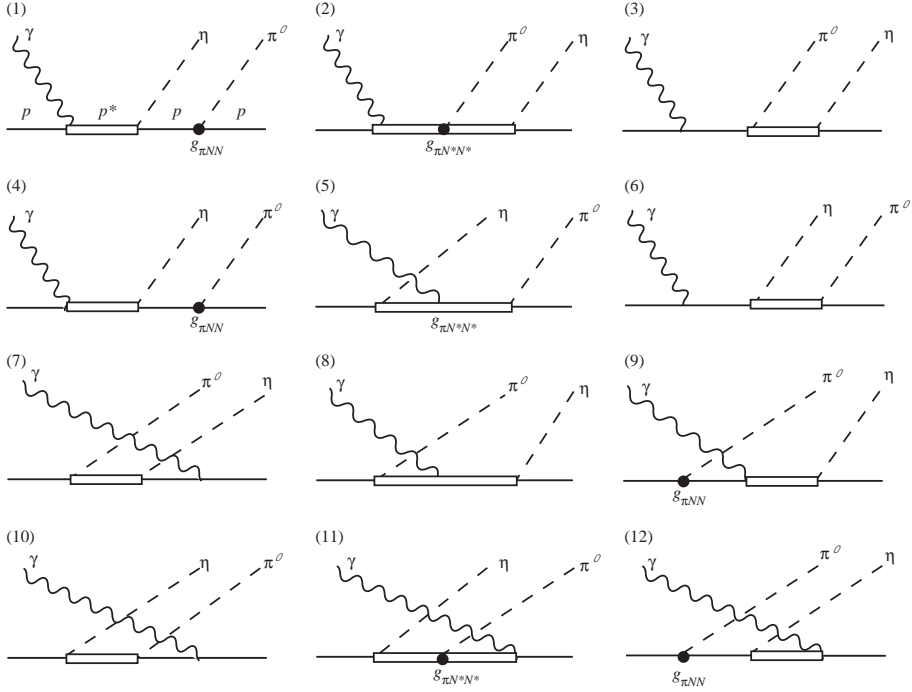}
    \begin{minipage}{12cm}
    \centering
    \caption{ \small
    Twelve $N^*$ dominant diagrams for $\gamma p \rightarrow
    \pi^0\eta p$ process.
    \label{12diag}}
    \end{minipage}
\end{figure}

Thus we compute the twelve resonance
dominant diagrams as shown in Figs.~\ref{12diag} in which
the initial pion is replaced by a photon.
For $\gamma NN$ vertex, we have
\begin{equation}
    \label{LgammaNN}
  {\cal L}_{\gamma NN}= - e \bar{N} \gamma_\mu {1 + \tau_3 \over 2} N
  A^\mu + {e \over 4 M} \bar{N} (\kappa_{S} + \kappa_{V} \tau_3)
  \sigma^{\mu\nu} N F_{\mu\nu} + {\rm h.c.}
\end{equation}
where the $\kappa_{S}$ and $\kappa_{V}$ are related to
the isoscalar and isovector anomalous magnetic moments and take the
values
\begin{equation}
   \kappa_S = -0.12 \, \; \; \; \kappa_V = 3.7 \, .
\end{equation}
For $\gamma N^*N^*$, we assume the same form as (\ref{LgammaNN})
but the nucleon mass is replaced by the resonance mass.
The unknown magnetic moment of $N^*$
($\equiv \kappa^{**}$) is taken just the same as that
of the nucleon.
The uncertainty from this is, however,
not very important, since the diagrams
which contain the $\gamma N^* N^*$ coupling
(Figs.~\ref{12diag} (5) and (8)) play only a minor role.
For $\gamma NN^*$ vertex, we take the tensor form which is compatible
with gauge invariance:
\begin{equation}
  {\cal L}_{\gamma NN^*} =
  { i e  \kappa_{V}^* \over 2 (m_{N^*} + m_N)} \bar{N}^*
  \tau_3 \gamma_5 \sigma^{\mu\nu} N
  F_{\mu\nu} + {\rm h.c.}
\end{equation}
where we used isovector dominance and the magnetic coupling is given
as
\begin{equation}
   \kappa_V^* = 0.9 \, ,
\end{equation}
which is determined from the analyses of
eta photoproduction~\cite{benmerrouche}.

We still
need to consider signs of coupling constants carefully, since
there are more coupling constants in the photon induced reaction than
in the pion induced reaction.
The twelve diagrams of  Fig.~\ref{12diag} are classified into four
groups according to the common coupling constants:
\beq
{\rm Group \; A} &\sim&
{\rm Figs.}\;  (1)\;  (4) \; (9) \; {\rm and}\;  (12)
\sim
g_{\gamma NN^*} g_{\pi NN} \nonumber \\
{\rm Group \; B} &\sim&
{\rm Figs.}\;  (2)\;  {\rm and}\;  (11)
\sim
g_{\gamma NN^*} g_{\pi N^* N^*} \nonumber \\
{\rm Group \; C} &\sim&
{\rm Figs.}\;  (3)\;  (6) \; (7) \; {\rm and}\;  (10)
\sim
g_{\gamma NN} g_{\pi NN^*} \nonumber \\
{\rm Group \; D} &\sim&
{\rm Figs.}\;  (5)\; {\rm and}\;  (8)
\sim
g_{\gamma N^*N^*} g_{\pi NN^*} \nonumber
\eeq
Here we do not include another coupling constant $g_{\eta NN^*}$
which is common to all the diagrams, since only relative signs
are our concern.
Also, the photon coupling constants $g_{\gamma NN}$ etc are symbolic,
since actual couplings contain both Dirac (vector) and Pauli (tensor)
types as given in (\ref{LgammaNN}).

\begin{figure}[tbp]
    \centering
    \footnotesize
    \epsfxsize = 7cm
    \epsfbox{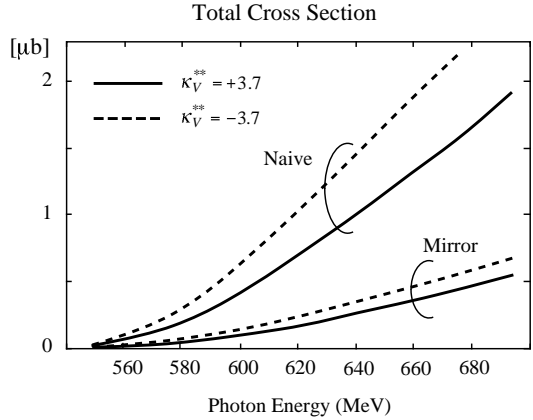}
    \begin{minipage}{14cm}
	\centering
	\caption{ \small
	Total cross sections for
	$\gamma p \to \eta \pi^0 p$ for the naive and mirror models as
	functions of the initial pion momentum
	$P_{c.m.}$ in the center of mass frame.  \label{gamtot}}
    \end{minipage}
\end{figure}

Since we are concerning  only about relative signs, let us
multiply $g_{\pi NN^*}$ to the above combinations:
\beq
(A) &\to& g_{\gamma NN^*} g_{\pi NN^*} g_{\pi NN} \; ,
\label{typeA} \\
(B) &\to& g_{\gamma NN^*} g_{\pi NN^*} g_{\pi N^* N^*} \; ,
\label{typeB} \\
(C) &\to& g_{\gamma NN} g_{\pi NN^*}^2 \; ,
\label{typeC} \\
(D) &\to& g_{\gamma N^*N^*} g_{\pi NN^*}^2 \; .
\label{typeD}
\eeq
The relative sign of $g_{\gamma NN^*}$ and $g_{\pi NN^*}$ is known
from the pion photoproduction;
it is plus for the proton and minus for the neutron~\cite{pdg2}.
The opposite signs for the proton and neutron is due to isovector nature
of the photon coupling to $N(1535)$.
Since we know the photon coupling to the nucleon, the sign ambiguity
among the groups (A), (B) and (C) is removed modulo the relative sign
of $g_{\pi NN}$ and $g_{\pi N^*N^*}$, which is the one under the
current interest.
The sign of the last group (D) can not be determined; it depends on
the sign of the coupling $g_{\gamma N^*N^*}$.
The Dirac coupling is determined since it is the charge of
$N(1535)$.
However, the strength of the
Pauli term can not be fixed; it depends on the unknown
magnetic moments of $N(1535)$.
It turns out, however, that the contribution of the
two terms of the group D, Figs.~\ref{12diag} (5) and (8), are
not very important.
We have verified this by computing cross sections with different values
of the anomalous magnetic moments of $N^*$.
Typical cases are shown in  Figs.~\ref{gamtot}
(see explanations below).
In this way, we can again discuss the relative sign of
$g_{\pi NN}$ and $g_{\pi N^*N^*}$ among the diagrams of the groups A,
B and C.

\begin{figure}[tbp]
    \vspace*{1cm}
    \centering
    \footnotesize
    \epsfxsize = 13cm
    \epsfbox{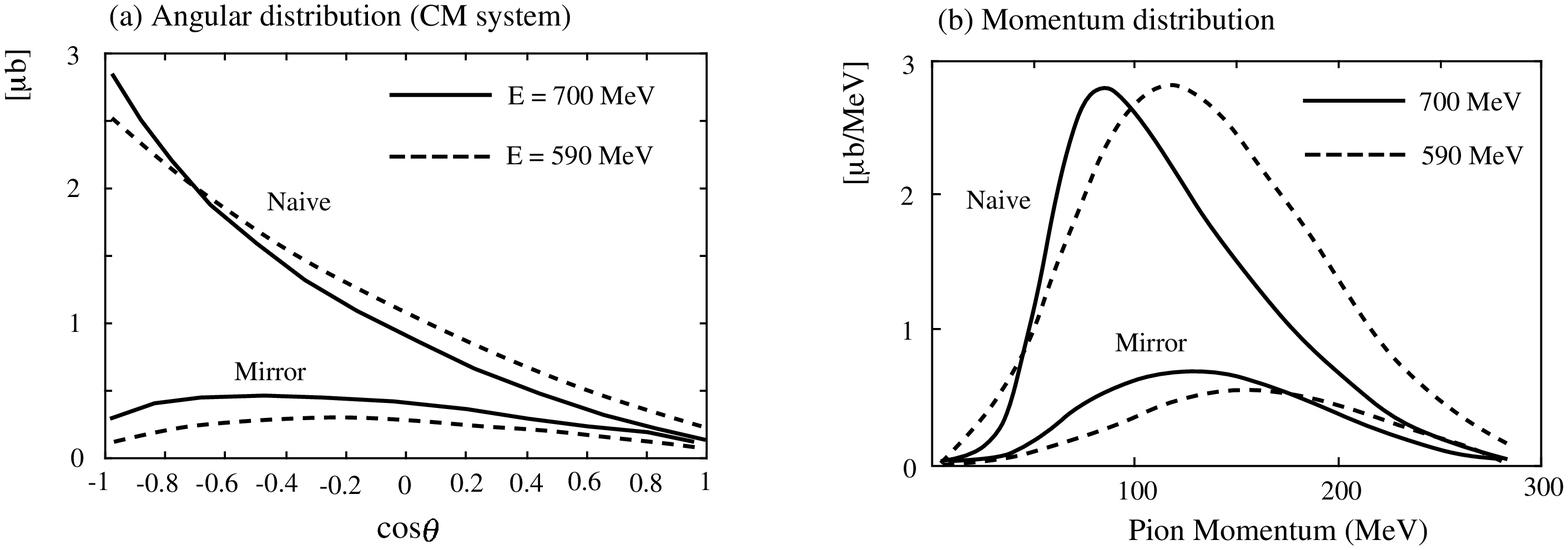}
    \begin{minipage}{14cm}
	\centering
	\caption{ \small
	(a) Angular and (b) momentum distributions of
	$\pi^0$ in the final state.
	Cross sections at the incident photon energy
	$E = 590$ MeV are scaled by 10 for the angular distributions and
	20/3 for the momentum distributions.
	Magnetic moments of $N^*$, $\kappa^{**}$, is taken to be the same as
	that of the nucleon.
	\label{gamang}}
    \end{minipage}
\end{figure}

As in the previous case, we have computed total cross sections as well
as momentum and angular distributions.
In Fig.~\ref{gamtot}, total cross sections are plotted as functions
of the incident photon energy.
Here we have shown the results using the anomalous magnetic moments
of $N^*$ which are the same as the nucleon and for those with opposite
sign.
The difference is within several ten percent, justifying relatively
less importance due to the ambiguity in these unknown couplings.
Unlike the pion induced reactions, the cross section of the naive
model is larger than the mirror model.
This is so because the dominant diagrams in the present process are
(1) and (2) of Figs.~\ref{12diag} and the approximate cancellation
between them are realized in the mirror model.
Accordingly, the angular and momentum distributions shown in
Figs.~\ref{gamang} follow this tendency in magnitude.
The shape of momentum distributions are qualitatively the same for the
naive and mirror models, while the angular distributions differ in
the two assignments,  reflecting once again the sign of the pion
p-wave coupling.

\section{Conclusion}

We have considered chiral symmetry aspects of baryons by assuming
that they belong to suitable linear representations of the
chiral symmetry group.
We have emphasized a possibility that the positive and negative parity
nucleons form a large representation of the chiral group where
the positive and negative parity
nucleons transform to each other under axial transformations.
This particular representation has been known for some time but we have
shown that this is one of two possible representations when there are
two nucleons.

A distinguished feature of the mirror representation is that the
different parity nucleons can form a degenerate multiplet with a
finite mass $m_{0}$ when chiral symmetry is restored.
This is very much different from the ordinary linear sigma model, in
which the nucleon masses are generated solely by spontaneous
breaking of chiral symmetry.
The origin of finite $m_{0} \neq 0$ is beyond
the dynamics of chiral symmetry.
It is interesting if more information can be extracted from
lattice QCD or QCD sum rules methods.

Another aspect of the mirror model is that the negative parity
nucleon carries the negative axial charge.
In effective theories, it is a group theoretical
consequence.
However, microscopic origin should be investigated, as baryons are
composite objects.
Further study of this is absolutely necessary~\cite{nacher}.

After establishing general features of the chiral symmetry
representations for baryons, we have studied microscopic origin
by using baryon interpolating fields.
It turns out that widely used interpolating fields of
three quarks possess the naive chiral property.
This can be seen both directly through commutation relations and
indirectly by computing the Yukawa pion coupling.
By contrast, the mirror chiral property is realized by five quark
operators.
This can be shown once again both in commutation relations and pion
couplings.

Finally, we have considered experiments to observe the
theoretical issues discussed here.
There is a fortunate situation that the negative parity resonance
$N(1535)$ can be seen almost selectively by detecting eta in the
final state.
Using this, we have proposed eta and pion productions in pion or
photon induced reactions.
interference effects are then used to observe the relative sign of
$g_{\pi NN}$ and $g_{\pi N^* N^*}$ which can distinguish either naive
or mirror property of the nucleon.

Although chiral symmetry has a long history in strong interaction
physics, its realization in the baryon sector has not been
investigated very much so far.
It serves an interesting subject in future hadron physics.


\section*{Acknowledgements}
We would like to thank Prof. T. Kunihiro, Prof. T. Hatsuda, Dr. H. Kim and
Dr. Y. Nemoto for discussions.  This work is supported
in part by the Grant-in-Aid for scientific research (C)(2)
11640261. The work by D.J. is supported by the
Spanish Ministry of Education in the Program ``Estancias de Doctores y
Tecn\'ologos Extranjeros en Espa\~{n}a''.
%

\end{document}